\documentclass[preprint]{aastex}
\usepackage{emulateapj5}
 
\newcommand{\zem}{{\ifmmode{z_{em}}\else{$z_{em}$}\fi}}
\newcommand{\zabs}{{\ifmmode{z_{abs}}\else{$z_{abs}$}\fi}}
\newcommand{\kms}{{\ifmmode{{\rm km~s}^{-1}}\else{km~s$^{-1}$}\fi}}
\newcommand{\delv}{{\ifmmode{\Delta v}\else{$\Delta v$}\fi}}
\newcommand{\cmm}{{\ifmmode{{\rm cm}^{-2}}\else{cm$^{-2}$}\fi}}
\newcommand{\cmmm}{{\ifmmode{{\rm cm}^{-3}}\else{cm$^{-3}$}\fi}}
\newcommand{\nhi}{{\ifmmode{N_{\rm H\;I}}\else{$N_{\rm H\;I}$}\fi}}

\newcommand{\lognhi}{$\log N_{\rm HI}$}
\newcommand{\lognmgii}{$\log N_{\rm MgII}$}
\def\lsim{\lower0.3em\hbox{$\,\buildrel <\over\sim\,$}}
\def\gsim{\lower0.3em\hbox{$\,\buildrel >\over\sim\,$}}

\newcounter{species} 
\def\ion#1#2{\setcounter{species}{#2}#1$\;${\scriptsize\Roman{species}}\relax}

\newcommand{\lya}{Ly$\alpha$}
\newcommand{\lyb}{Ly$\beta$}
\newcommand{\hi}{{\rm H}~{\sc i}}

\slugcomment{\today}
\shorttitle{Super-Solar Metallicity in Weak \ion{Mg}{2} Systems}
\shortauthors{Misawa et al.}

\begin{document}

\title{SUPER-SOLAR METALLICITY IN WEAK \ion{Mg}{2} ABSORPTION SYSTEMS
       AT $z$ $\sim$ 1.7\altaffilmark{1}}

\altaffiltext{1}{Based on archive data of observations made with the
                 ESO Telescopes at the La Silla or Paranal
                 Observatories under programs, 65.O-0411, 66.A-0221,
                 67.A-0280}

\author{Toru Misawa\altaffilmark{2},
        Jane C. Charlton\altaffilmark{2}, and
        Anand Narayanan\altaffilmark{2}}

\altaffiltext{2}{Department of Astronomy \& Astrophysics, The
  Pennsylvania State University, University Park, PA 16802}

\email{misawa, charlton, anand@astro.psu.edu}

\begin{abstract}

Through photoionization modeling, constraints on the physical
conditions of three $z$ $\sim$ 1.7 single-cloud weak \ion{Mg}{2}
systems ($W_r$(2796) $\leq$ 0.3\AA) are derived.  Constraints are
provided by high resolution $R$ $=$ 45,000, high signal-to-noise
spectra of the three quasars HE~0141$-$3932, HE~0429$-$4091, and
HE~2243$-$6031 which we have obtained from the European Southern
Observatory (ESO) archive of Very Large Telescope (VLT), Ultraviolet
and Visual Echelle Spectrograph (UVES). Results are as follows.

\begin{description}
 
 \item[(1)] The single-cloud weak \ion{Mg}{2} absorption in the three
 $z$ $\sim$ 1.7 systems is produced by clouds with ionization
 parameters of $-$3.8 $<$ $\log U$ \lsim\ $-$2.0 and sizes of 1 --
 100~pc.

 \item[(2)] In addition to the low-ionization phase \ion{Mg}{2}
 clouds, all systems need an additional 1 -- 3 high-ionization phase
 \ion{C}{4} clouds within 100 \kms\ of the \ion{Mg}{2} component.  The
 ionization parameters of the \ion{C}{4} phases range from $-$1.9 $<$
 $\log U$ $<$ $-$1.0, with sizes of tens of parsecs to kiloparsecs.
 In all cases at least one \ion{C}{4} cloud is centered at the same
 velocity (within 5 \kms\ in our systems) as the \ion{Mg}{2} clouds.

 \item[(3)] Two of the three single-cloud weak \ion{Mg}{2} absorbers
 have near-solar or super-solar metallicities, if we assume a solar
 abundance pattern. Although such large metallicities have been found
 for $z$ $<$ 1 weak \ion{Mg}{2} absorbers, these are the first high
 metallicities derived for such systems at higher redshifts.  These
 strong constraints were possible because of the specific shapes of
 the \lya\ profiles in these cases.  All single-cloud weak \ion{Mg}{2}
 absorbers may have high metallicities, but in some cases the
 kinematic spread of the \ion{C}{4} cloud contributions to \lya\ do
 not allow such a determination.

 \item[(4)] Two of the three weak \ion{Mg}{2} systems also need
 additional low-metallicity, broad \lya\ absorption lines, offset in
 velocity from the metal-line absorption, in order to reproduce the
 full \lya\ profile.

 \item[(5)] Metallicity in single-cloud weak \ion{Mg}{2} systems are
 more than an order of magnitude larger than those in Damped \lya\
 systems at $z$ $\sim$ 1.7.  In fact, there appears to be a gradual
 decrease in metallicity with increasing \nhi, from these, the most
 metal-rich \lya\ forest clouds, to Lyman limit systems, to sub-DLAs,
 and finally to the DLAs. Weak \ion{Mg}{2} absorbers could be near
 local pockets in which star-formation has occurred, but where there
 is little gas to dilute the metals that are dispersed into the
 region, resulting in their very high metallicities.  We speculate
 that DLAs may be subject to the opposite effect, where a large
 dilution of metals produced in the vicinity will occur, leading to a
 small metallicity.
 
 \end{description}
\end{abstract}


\section{Introduction}
\label{sec:intro}

Many single-cloud weak \ion{Mg}{2} absorbers ($W_r$(2796) $\leq$
0.3\AA) at 0.4 $<$ $z$ $<$ 1.0 have been found to have solar or even
super-solar metallicities \citep{rig02,cha03}. This is quite puzzling
in view of the fact that they tend to be found at distances of 40 --
100~kpc from star-forming galaxies \citep{chu05,mil06}, and not near
detected sites of star formation.  These objects have a cross-section
similar to the absorption cross section of luminous galaxies, thus
they occupy a significant volume of the universe \citep{rig02}.
Because weak \ion{Mg}{2} systems only contain a small mass, they do
not trace the cosmic metal abundance, however, it is still important
to understand local star-formation activities in the full range of
environments over cosmic time.

\citet{nar07} presented the results of a search for weak \ion{Mg}{2}
absorption in VLT/UVES quasar spectra over a cumulative redshift path
of 77.3 at 0.4 $<$ $z$ $<$ 2.4.  A total of 116 \ion{Mg}{2} absorbers
with 0.02 $<$ $W_r(2796)$ $<$ 0.3~\AA\ were detected, with $\sim$ 60\%
of these having just a single-cloud, narrow component in \ion{Mg}{2}.
It has been argued that a large fraction of the population of
multiple-cloud weak \ion{Mg}{2} absorbers (i.e., absorbers with
multiple-clouds, which are resolved by high-resolutional ($R$ $\sim$
40,000) spectroscopies) have a different physical origin from the
single-cloud weak \ion{Mg}{2} absorbers \citep{rig02,mas05,din05}.
\citet{nar07} found that the redshift path density, $dN/dz$, of weak
\ion{Mg}{2} absorbers increases from $z$ $=$ 2.4 to $z$ $=$ 1.2, where
it peaks, and subsequently declines until the present.
  
The paper of \citet{nar07} followed a weak \ion{Mg}{2} survey of
\citet{lyn06}, which used a small subset of high quality VLT/UVES
quasar spectra.  From that survey, \citet{lyn07} selected the nine $z$
$>$ 1.4 weak \ion{Mg}{2} absorbers and applied photo-ionization
models.  Of these, four were single-cloud weak \ion{Mg}{2} absorbers
with coverage of the corresponding \lya\ transitions so that
metallicities could be constrained.

In general, \citet{lyn07} found that the physical conditions of the
$z$ $>$ 1.4 weak \ion{Mg}{2} absorbers were similar to those of the
same class of absorbers at lower redshift, suggesting that the same
production mechanisms are responsible over this time period.  However,
in contrast to the situation at low redshift \citep{rig02,cha03},
\citet{lyn07} were only able to derive lower limits to the
metallicity, ranging from 1/100$^{th}$ -- 1/10$^{th}$ of the solar
value.  Single-cloud weak \ion{Mg}{2} absorbers, in addition to the
cloud that produced the \ion{Mg}{2} absorption, have separate, lower
density \ion{C}{4} clouds that are spread over tens to hundreds of
\kms\ \citep{cha03,lyn07}.  The reason that only lower limits could be
derived for the metallicity in these cases was because of the
contributions to the \lya\ profile from the kinematically spread
\ion{C}{4} clouds.  On the other hand, \citet{lyn07} also suggested
that there could be a build-up of metals in the population of
single-cloud weak \ion{Mg}{2} absorbers from $z$ $\sim$ 2 to $z$
$\sim$ 0.4.

In view of the large spread of metallicity constraints at any
redshift, a sample of only four $z$ $>$ 1.4 single-cloud weak
\ion{Mg}{2} absorbers was not sufficient to resolve the issue of
whether metallicity evolution occurs in the population.  We therefore
identified three additional $z$ $>$ 1.4 candidates for
photo-ionization modeling from the larger survey of \citet{nar07}.
These systems, the $z$ $=$ 1.78169 absorber toward HE~0141$-$3932, the
$z$ $=$ 1.68079 absorber toward HE~0429$-$4091, and the $z$ $=$
1.75570 absorber toward HE~2243$-$6031, have high $S/N$-ratio VLT/UVES
coverage of key constraining metal-line transitions, as well as \lya.
We apply Cloudy \citep{fer98} photoionization models to constrain the
ionization parameters and metallicities of the phases\footnote{Phases
are defined as regions of gas with similar metallicity, temperature,
and volume density.} of gas that are required to fit the data.  Our
focus is on the question of the metallicity evolution of the
single-cloud weak \ion{Mg}{2} absorbers, and on comparing the result
to the metallicity evolution of other classes of quasar absorption
line systems.

We will begin in \S~\ref{sec:data} with a summary of the data that we
have used to constrain the properties of the three $z$ $\sim$ 1.7
single-cloud weak \ion{Mg}{2} absorbers.  Next, in \S~\ref{sec:photo},
we describe our procedure for photo-ionization modeling with Cloudy,
which is based on comparing synthetic model profiles to the data.
\S~\ref{sec:systems} presents the detailed results for the three
absorbers.  Finally, in \S~\ref{sec:discussion}, we summarize those
results, compare the metallicities of single-cloud weak \ion{Mg}{2}
absorbers to DLAs, to sub-DLAs, and to Lyman limit absorbers, and
discuss the implications of this comparison.  Throughout this paper,
we use a cosmology with $H_{0}$=72 \kms Mpc$^{-1}$, $\Omega_{m}$=0.3,
and $\Omega_{\Lambda}$=0.7.

\section{Data}
\label{sec:data}

All three weak \ion{Mg}{2} systems in our sample were found in the
survey of \citet{nar07} who studied the statistical properties of weak
\ion{Mg}{2} systems, using 81 VLT/UVES spectra.  Data were retrieved
from the ESO archive, and reduced using the ESO MIDAS pipeline. In the
case of multiple epoch observations, the individual exposures were
combined to enhance the $S/N$ ratio.  The detailed description of the
original data analysis is presented in \citet{nar07}.

The main purpose of this paper is to place strict constraints on
physical properties (such as metallicity, ionization
parameter\footnote{The ionization parameter is defined the ratio of
ionizing photons ($n_{\gamma}$) to the number density of hydrogen in
the absorbing gas ($n_{H}$).}, gas temperature, and size) of single
cloud, weak \ion{Mg}{2} absorbers at high redshift. Among the 116 weak
\ion{Mg}{2} systems in \citet{nar07}, 16 systems at relatively high
redshift ($z$ $>$ 1.5) have \lya\ absorption lines covered with the
observed spectra. \lya\ is quite important for determining
metallicities. Of the 16 systems, 7 are classified as single-cloud
systems. Four of these seven, the $z$ = 1.65146 system toward
HE0001$-$2340, the $z$ = 1.70849 system toward HE0151$-$4326, the $z$
= 1.79624 system toward HE2347$-$4342, and the $z$ = 2.17455 system
toward HE0940$-$1050 had already been modeled as described in
\citet{lyn07}, since they were first identified in the earlier survey
of \citet{lyn06}.  The other three systems, the $z$ = 1.78169 system
toward HE~0141$-$3932, the $z$ = 1.68079 system toward HE~0429$-$4091,
and the $z$ = 1.75570 system toward HE~2243$-$6031, are modeled in the
present paper.

In Table~\ref{tab1}, we give an observation log for the three quasars
toward which the single-cloud weak \ion{Mg}{2} absorbers were
detected.  Column (1) is the quasar name, columns (2) and (3) are the
coordinates of the quasars, columns (4) and (5) are the emission
redshift and optical magnitude of the quasars, column (6) is observed
wavelength range, column (7) is the central wavelength for the blue
and red CCDs of VLT/UVES, columns (8) and (9) are total exposure time
and observing date, and column (10) is the proposal ID.  By chance,
all three quasars were observed as part of programs with the same
P.I., Sebastian Lopez.

The absorption profiles for key transitions used to constrain the
three single-cloud weak \ion{Mg}{2} absorbers are displayed in Figures
\ref{fig1}, \ref{fig2}, and \ref{fig3}.  Rest-frame equivalent widths
and $5\sigma$ rest-frame equivalent width limits for these same
transitions are listed in Table~\ref{tab2}.

\section{Photoionization Models}
\label{sec:photo}

In this section we briefly summarize the strategy for photoionization
modeling, which is similar to the procedure practiced in previous
studies \citep[e.g.,][]{chu99,cha03,lyn07}.  At first, we determine
the line parameters (column density, Doppler parameter, and redshift)
for the \ion{Mg}{2} doublets, using a Voigt profile fitting code ({\sc
MINFIT}; \citealt{chu03}).  For each of the three absorbers modeled
here, a single absorption component provided an adequate fit to the
\ion{Mg}{2} doublet.  The best fit parameters are listed in
Table~\ref{tab3}: column (1) is the name of the transition, columns
(2), (3), and (4) are the fit parameters of velocity shift from the
system central redshift\footnote{The system central redshift is
defined as the flux-weighted center of the \ion{Mg}{2}~$\lambda$2796
absorption line.}, column density, and Doppler parameter with their
1$\sigma$ errors.  For completeness, we also applied formal Voigt
profile fits to other detected transitions, and listed the results in
Table~\ref{tab3}.  We emphasize, however, that we did not use these
values (except for the optimized transitions) to determine model
constraints.  Instead we compared directly to the shapes of the
observed absorption profiles as described below.

Assuming that the absorbing clouds are in photoionization equilibrium,
their ionization conditions are derived using the photoionization code
Cloudy, version 07.02.00 \citep{fer98}, optimizing on the observed
column density of \ion{Mg}{2} and comparing model predictions to other
observed low ionization transitions. The clouds in each phase are
modeled as plane-parallel structures of constant density.  The
ionization parameter ($\log U$ $=$ $\log[n_{\gamma}/n_H]$) and
metallicity ($Z/Z_{\odot}$) of the clouds serve as free parameters in
the modeling. The elemental abundance pattern is initially assumed to
be solar, and variations are explored when they are suggested by the
data.

Following \citet{haa96,haa01}, we take the combined flux from quasars
and star forming galaxies at $z \sim 1.7$ (with a photon escape
fraction of 0.1) as an extragalactic background radiation (i.e.,
incident flux on the absorbers). We will consider the possible effects
of a nearby stellar contribution to the ionizing radiation field in
\S~\ref{sec:assumptions}, although it should be safe to ignore them,
since weak \ion{Mg}{2} absorbers are not often within 40~kpc of bright
galaxies \citep{rig02,chu05}.

The metallicity and ionization parameter are initially varied in steps
of 0.5 dex, and then fine-tuned to select the model that corresponds
best with the observed data in steps of 0.1 dex. The column densities
for various ionization stages of each element and the equilibrium gas
temperature are provided by the Cloudy model.  The Doppler parameter
for each element can then be calculated from the expression for total
line width, $b_{tot}^{2}$ $=$ $b_{ther}^{2}$ $+$ $b_{turb}^{2}$, where
$b_{ther}$ ($=$ $\sqrt{2kT/m}$) is the thermal contribution to the
line width corresponding to a gas temperature $T$ and $b_{turb}$, the
contribution from internal gas turbulence. This latter quantity, which
is uniform across elements, is estimated using the gas temperature
from the Cloudy model and the observed Doppler parameter of the
element on which the model is optimized. From the derived column
densities and Doppler parameters, a synthetic spectrum is generated
and compared to the observed spectrum after convolving with a Gaussian
instrumental spread function ($R$ $=$ 45,000 in the case of our
VLT/UVES spectra).

Although the observed and the synthesized spectra are compared ``by
eye'', the method generally gives good results in the sense that other
models whose metallicity and ionization parameter are (sometimes only
slightly) different than the best values would make the model spectrum
deviate greatly from the observed spectrum.  It is not practical to
apply formal procedures to assess the quality of fit because of the
effect of blends and data defects on a $\chi^2$ statistic.  If all
transitions of a certain element are over/under-produced compared to
the observed spectrum in the best model, we consider a specific
abundance pattern for the system.  Our method is illustrated for a
specific example, the $v=1$~{\kms} cloud for the $z=1.78169$ system
toward HE0141-3932 (System 1) in Figure~\ref{fig4}.  We show, for this
component, that only values within the range $-0.7 < \log Z < 0.1$ and
within 0.1~dex of $\log U = -2.3$ produce acceptable fits to \lya,
\ion{Si}{2}, and \ion{Si}{4}, which are the main constraints for this
case.

Usually, low-ionization phase clouds that produce the \ion{Mg}{2}
absorption lines also produce other low-ionization transitions such as
\ion{Fe}{2}, \ion{Si}{2}, \ion{C}{2}, and \ion{Al}{2}. However, the
higher-ionization transition, \ion{C}{4}, and sometimes
intermediate-ionization transitions such as \ion{Si}{4} are not fully
produced by the \ion{Mg}{2} clouds. Therefore, once we find the best
model for the \ion{Mg}{2} clouds, we repeat a similar Cloudy analysis
for the \ion{C}{4} clouds. For the high-ionization phase, multiple
components are sometimes required to fit \ion{C}{4} absorption lines
detected at different velocities.  The results of Voigt profile fits
to these \ion{C}{4} components are also listed in Table~\ref{tab3}.
Finally, the combined models, with contributions from both low- and
high-ionization phases, are compared to the observed spectrum. Since
\lya\ absorption lines have contributions from both the low- and
high-ionization phases, determination of metallicities for the low-
and high-ionization phases can be degenerate.  However, in the three
single-cloud weak \ion{Mg}{2} absorbers that we are considering, it is
possible to place limits on metallicity of both phases, so as not to
over-produce the \lya\ absorption.  Also, in some cases, models with
high metallicity (leading to low temperatures) can be favored when
comparing model profiles to some observed low ionization transitions.

\section{Individual Systems}
\label{sec:systems}

For photoionization models, our procedure is adjusted for each system
based on the available constraints in various transitions.  Some
transitions are blended with other lines, and others may be
saturated. These issues are considered carefully during
photoionization modeling. In this section, we present the spectra of
our three weak \ion{Mg}{2} systems and then describe the results of
photoionization models. We can place strong constraints on metallicity
and ionization parameter because multiple transitions (including \lya)
have been detected in high S/N-ratio regions of the spectra (see
Figures~\ref{fig1} -- \ref{fig3}). We will, hereafter, refer to the
three weak \ion{Mg}{2} systems as Systems~1, 2, and 3.  After
presenting the constraints, we discuss in \S~\ref{sec:assumptions} the
effect of varying model parameters and assumptions such as
photoionization equilibrium, the shape of the incident radiation
field, and the abundance of dust.

\subsection{System~1 (HE~0141$-$3932; $z$=1.78169)}
\label{sec:sys1}

\begin{figure}
 \begin{center}
  \includegraphics[width=13cm,angle=0]{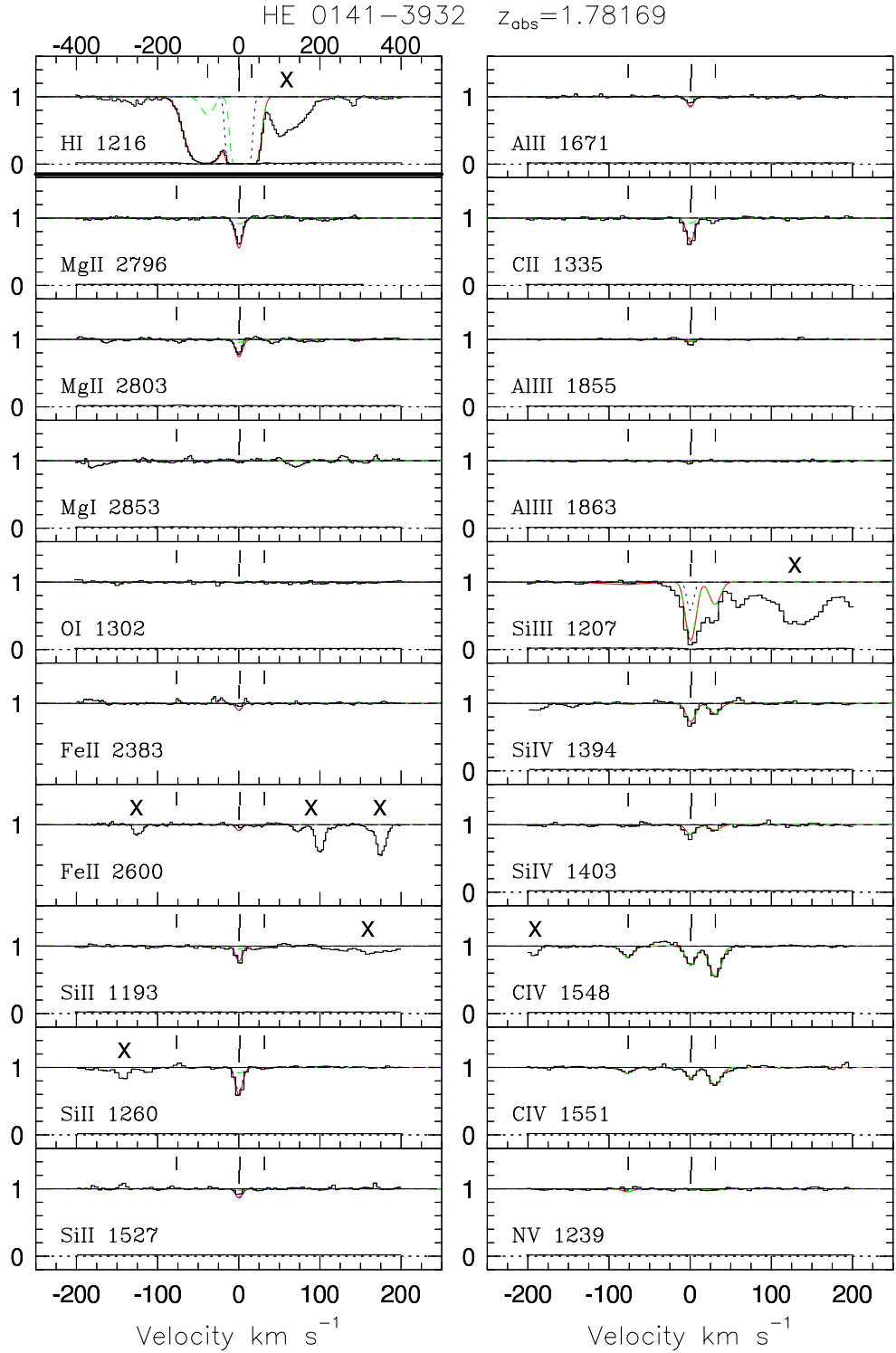}
 \end{center}
 \caption{Detected transitions, and those that provide limiting
          constraints, shown in velocity space for the single-cloud
          weak \ion{Mg}{2} system at $z$ = 1.78169 toward
          HE~0141$-$3932. The velocity range is $\pm$200~\kms\ for all
          transitions except for \lya\ whose range is $\pm$400~\kms.
          Vertical scale is from $-$0.2 to 1.6 except for
          \ion{Fe}{2}~$\lambda$2383 and \ion{Fe}{2}~$\lambda$2600 for
          which it is slightly zoomed from 0.2 to 1.4. The data are
          from a VLT/UVES spectrum with a resolution of $R$ =
          45,000. The sigma spectrum is indicated as a solid histogram
          just above the dotted line crossing each plot at zero
          flux. Features from un-related systems are marked with an
          "X". An example of the best model fit to the observed
          spectrum (summarized in Table~\ref{tab6}) is superimposed on
          the data as a solid (red) curve.  Contributions of the
          \ion{Mg}{2} and \ion{C}{4} clouds to the model are displayed
          separately as dotted (blue) and long-dashed (green) lines. A
          single tick centered on the \ion{Mg}{2} profile marks the
          position of the low ionization phase, while ticks in the row
          above mark the positions of the high ionization phase
          clouds.\label{fig1}}
\end{figure}

This system is detected in \lya, \ion{Mg}{2}, \ion{Si}{2},
\ion{Al}{2}, \ion{C}{2}, \ion{Al}{3}, \ion{Si}{3}, \ion{Si}{4}, and
\ion{C}{4}, as shown in Figure~\ref{fig1}.  The observed spectrum
covers \ion{Mg}{1}, \ion{O}{1}, and \ion{N}{5}, but these transitions
are not detected to a 5$\sigma$ detection limit.  Although
\ion{Fe}{2}~$\lambda$2383 is formally detected at only 3$\sigma$ and
may be contaminated by spurious features, \ion{Fe}{2}~$\lambda$2600 is
also detected at 2.5$\sigma$ and both are precisely aligned with the
\ion{Mg}{2}.  We therefore are confident in applying the \ion{Fe}{2}
as a constraint.  Rest-frame equivalent widths and $5\sigma$
equivalent width limits are listed in Table~\ref{tab2}. Although the
\lya\ profile is complex, it provides a very strong constraint on
metallicity because the flux at the position of the component centered
on \ion{Mg}{2} begins to recover on its blue side, which enables us to
fit this component of \lya\ effectively. All transitions except for
high-ionization lines such as \ion{C}{4} and \ion{Si}{4} have only
single components detected.

We begin by fitting a Voigt profile to the \ion{Mg}{2} doublet.  The
best fit parameters are listed in Table~\ref{tab3}.  A lower limit to
the ionization parameter is provided by \ion{Fe}{2}, which is
over-produced at $\log U$ $<$ $-$3.8.  \ion{Mg}{1} is also
over-produced at $\log U$ $<$ $-$6.0.  (\ion{Mg}{1} was not detected
in the observed spectrum, and \ion{Fe}{2} is only detected in the
$\lambda$2383 transition at a level of $\sim$ 3 $\sigma$.)  There is
also a strict upper limit on the ionization parameter of $\log U$ $<$
$-$2.8 above which the high-ionization transitions \ion{C}{4} and
\ion{Si}{4} are over-produced.  Based on these considerations, models
with $\log U$ $=$ $-$3.8 -- $-$2.8 seem acceptable, however
\ion{Si}{2} and \ion{Al}{2} provide additional constraints.  An
ionization parameter, $-$3.8 $<$ $\log U$ $<$ $-$3.7, is consistent
with all constraints.  At lower values of $\log U$ \ion{Fe}{2} is
over-produced, and at higher values \ion{Al}{2} and \ion{Si}{2} are
over-produced.  Independent of ionization parameter, the metallicity
must be $\log Z$ $>$ $-$0.7 in order that \lya\ is not over-produced.
An upper limit of $\log Z$ $<$ $-$0.5 would also apply, for solar
abundance pattern, if $\log U$ $=$ $-$3.7, in order that \ion{Fe}{2}
absorption is not over-produced.  The lower limit on metallicity of
$\log Z$ $>$ $-$0.7 applies for the solar abundance pattern, and would
be adjusted downwards for an $\alpha$-enhanced pattern.  Similarly,
$\log U$ could be lower in the $\alpha$-enhanced case.  If dust
depletion was important, the lower limit on metallicity would be
increased.

In addition to the \ion{Mg}{2} cloud, three more higher-ionization
clouds are needed to reproduce the three detected \ion{C}{4}
components, the two \ion{Si}{4} components, and the right wing of the
\lya\ profile. There are no \ion{N}{5} lines detected, which is useful
to place an upper limit on the ionization parameter.  The results of
Voigt profile fits to the \ion{C}{4} doublet are given in
Table~\ref{tab3}.  Beginning with these \ion{C}{4} fits, we adjust the
ionization parameter and metallicity to satisfy constraints from other
transitions.  In order to avoid over-production of \ion{Si}{4} and low
ionization transitions (at low values) or \ion{N}{5} (at high values),
the ionization parameters for the additional 3 clouds at \delv\ $=$
$-$77, 1, and 31~\kms\ from the system center should be $-$1.5 $<$
$\log U$ $<$ $-$1.0, $-$2.35 $<$ $\log U$ $<$ $-$2.25, and $-$1.9 $<$
$\log U$ $<$ $-$1.8, respectively.  For these ionization parameters,
lower limits on metallicity of $\log Z$ $>$ $-1.8$, $\log Z$ $>$
$-0.7$, and $\log Z$ $>$ $-0.5$ will apply for the three \ion{C}{4}
components in order that \lya\ is not over-produced.  For the first
two components, at \delv\ $=$ $-$77 and 1~{\kms} upper limits on $\log
Z$ of 0.5 and 0.1 in order that low ionization transitions are not
over-produced.  The third \ion{C}{4} component, at \delv\ $=$ 31~\kms
, has a strong constraint on its metallicity, assuming that it is
responsible for the \lya\ absorption in the red wing of the profile
(since the \ion{Mg}{2} cloud cannot account for that absorption
without over-producing \lya\ in the blue wing).  For that \ion{C}{4}
we find that $-$0.5 \lsim\ $\log Z$ \lsim\ $-$0.4 in order to match
the \lya\ profile.  We emphasize that the three components have
different ionization parameters from each other, although these values
are all higher than that of the \ion{Mg}{2} cloud.

The one \ion{Mg}{2} and three \ion{C}{4} clouds reproduced all
transitions except for \lya. There are absorption features at both
sides of the \lya\ profile, at \delv\ $<$ $-$50 and at \delv\ $>$
70~\kms\ (see Figure~\ref{fig1}). Neither of them are \lyb\ profiles
from higher redshift systems, because the corresponding \lya\
absorption lines would be located redward of the quasar \lya\ emission
line.  The \ion{C}{4} component at \delv\ = $-$77~\kms\ cannot itself
give rise to all of the observed \lya\ around that velocity because it
is too narrow.  Thus, we fit the broad \lya\ profile ($b(H)$ $=$
42~\kms\ as listed in Table~\ref{tab3}), and placed upper limits on
its metallicity: ($\log Z$ $<$ 0.0 for $\log U$ $\sim$ $-$3 and $\log
Z$ $<$ $-$2.0 for $\log U$ $\sim$ $-$1.5) so as not to produce other
transitions at this velocity.  Ionization parameter is not constrained
for this broad \lya\ cloud, although high values of $\log U$ are ruled
out unless the metallicity is extremely small.

We conclude that, for a solar abundance pattern, the metallicity of
the $z$ $=$ 1.78169 weak \ion{Mg}{2} system in the HE~0141-3932
spectrum is constrained to be $-$0.7 \lsim\ $\log Z$ \lsim\ $-$0.5.  A
strict lower limit on metallicity of $\log Z$ $>$ $-$0.7 applies in
order that \lya\ absorption is not over-produced.  Two of the three
\ion{C}{4} clouds related to this system are constrained to have
similar or higher metallicities, and for the other no significant
metallicity constraint is available.  The \ion{Mg}{2} cloud has a
sub-parsec size, much smaller than the three \ion{C}{4} phase
absorbers, with sizes of $>$10~pc. All components are optically thin,
the \ion{Mg}{2} cloud giving rise to \lognhi\ $\sim$ 15.9.

A summary of how specific transitions were used to constrain $\log U$
and $\log Z$ for this system is given in Table~\ref{tab4}.  Ranges of
acceptable model parameters for each model cloud are listed in
Table~\ref{tab5}, in which column (1) is the absorption redshift,
column (2) is the optimizing transition used in Cloudy, column (3) is
the velocity shift from the system center, and column (4) is the
acceptable range of ionization parameter.  Column (5) of the table
presents the best constraint on metallicity, considering both
comparison to the observed \lya\ and the fit to metal-line
transitions, assuming a solar abundance pattern.  Column (6) is a more
conservative constraint on the metallicity of the cloud, using only
the requirement that \lya\ absorption is not over-produced.  Column
(7) is the depth of absorber along the line of sight, column (8) is
the Doppler parameter of the optimizing transition (listed in column
(2)), and column (9) is the column density of neutral hydrogen gas in
the absorbers.  An example of a best fit model and its specific
parameters is presented in Figure~\ref{fig1} and Table~\ref{tab6}.

\subsection{System~2 (HE~0429$-$4091; $z$=1.68079)}
\label{sec:sys2}

\begin{figure}
 \begin{center}
  \includegraphics[width=13cm,angle=0]{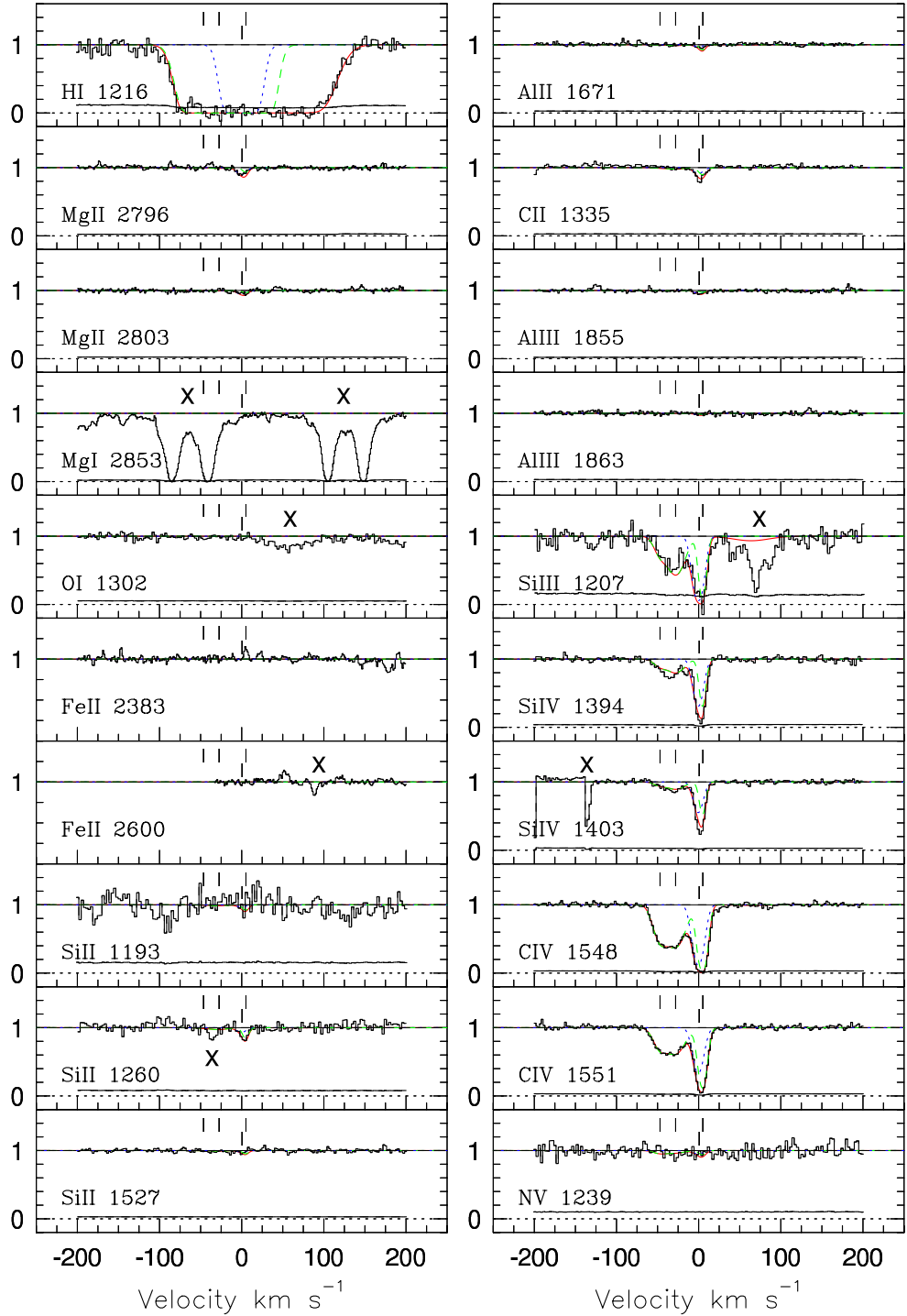}
 \end{center}
 \caption{Same as Figure~\ref{fig1}, but for the system at $z$ = 1.68079
          toward HE~0429$-$4901.\label{fig2}}
\end{figure}

This is a very weak \ion{Mg}{2} system whose rest-frame equivalent
width ($W_{r}$(2796) = 0.015\AA) is just above the lower limit of the
survey presented in \citet{nar07}. In this system, \lya, \ion{Si}{2},
\ion{Al}{2}, \ion{C}{2}, \ion{Al}{3}, \ion{Si}{3}, \ion{Si}{4}, and
\ion{C}{4} are clearly detected, but \ion{O}{1}, \ion{Fe}{2}, and
\ion{N}{5} are not detected at 5$\sigma$, as shown in
Figure~\ref{fig2}.  The rest-frame equivalent widths and $5\sigma$
limits are tabulated in Table~\ref{tab2}.

Again, we begin by fitting Voigt profiles to the \ion{Mg}{2} doublet
(see Table~\ref{tab3}).  Low ionization parameters, smaller than $\log
U$ $=$ $-$3.5, are not acceptable because of over-production of
\ion{Fe}{2}. At $\log U$ $>$ $-$2.0, \ion{C}{4} is over-produced.  For
this system, the \lya\ absorption is strong, and it probably arises
from offset \ion{C}{4} components rather than from the \ion{Mg}{2}
cloud.  Because of the strong \lya\ absorption we cannot place a
meaningful constraint on the metallicity of the system; the blue wing
of \lya\ only requires $\log Z$ $>$ $-$3.0.  However, at $-$3.5 $<$
$\log U$ $<$ $-$2.0, \ion{Al}{2} is over-produced at low metallicity
(such that $\log Z$ $>$ $-$1.0 for $\log U \sim$ $-$3.0 and $\log Z$
$>$ $-$0.3 for $\log U \sim$ $-$2.0).  Because the acceptable range of
ionization parameter is large for this system, we consider two extreme
cases: (i) the \ion{Mg}{2} cloud has the highest acceptable ionization
parameter ($\log U$ $=$ $-$2.0) that produces the maximum possible
\ion{C}{4} and \ion{Si}{4} absorption, and (ii) the \ion{Mg}{2} cloud
has the lowest acceptable ionization parameter ($\log U$ $=$ $-$3.5)
so that it does not significantly contribute to the \ion{C}{4} and
\ion{Si}{4} profiles. We will refer to these below as the maximal and
minimal \ion{C}{4} models.

In the maximal \ion{C}{4} model, we need an additional three
high-ionization components to fit \ion{C}{4} profile at \delv\ $=$
$-$47, $-$28, and 5~\kms\ from the system center. The fitting
parameters, after removing the contributions to \ion{C}{4} from the
low-ionization clouds, are listed in Table~\ref{tab3}.
Figure~\ref{fig2} also shows that the \ion{Mg}{2} cloud at 0~\kms\ is
offset and is not broad enough to explain all of the \ion{C}{4}
absorption at a similar velocity.  Only the \ion{C}{4} cloud at \delv\
$=$ $-$47~\kms\ has its metallicity constrained to be $\log Z$ $=$
$-$1.1 -- $-$1.0, assuming that it accounts for the blueward edge of
the saturated \lya\ profile.  The metallicity cannot be lower than
this value or \lya\ would be overproduced.  However, a constraint on
the ionization parameters of the three \ion{C}{4} clouds can be
obtained by the requirement that they also fit the \ion{Si}{3} and
\ion{Si}{4} doublet profile.  For the three \ion{C}{4} clouds, we find
$\log U$ in the range between $-$1.8 and $-$1.6.  The only
disagreement between this maximal \ion{C}{4} model and the observed
spectrum is an over-production of \ion{N}{5}. Adjusting the abundance
pattern so that nitrogen is reduced by at least 0.5 dex compared to
the other elements gives adequate results (see Figure~\ref{fig2}).  In
fact, the production mechanisms for nitrogen are poorly understood
\citep[e.g.,][and references therein]{rus92}. Namely, it is not well
understood yet whether massive stars or intermediate-mass stars (or
both) contribute to the {\it primary} nitrogen production before the
{\it secondary} production starts through the CNO cycle
\citep{spi05}. This nitrogen ``problem'' has also been reported
frequently in DLA systems \citep[e.g.,][]{pet02} and for the
Magellanic Bridge \citep[e.g.,][]{leh01}, as well as for
multiple-cloud weak \ion{Mg}{2} systems \citep{zon04}. Once the best
model parameters for the three \ion{C}{4} components are found, we
adjust the parameters for \ion{Mg}{2} again to compensate, and get
$\log U$ $=$ $-$2.0 and $\log Z$ $=$ $-$0.2 -- 0.1. The lower limit on
metallicity was placed in order not to over-produce \ion{Al}{2}, and
the upper limit in order not to under-produce \ion{Si}{4} and
\ion{C}{4}.  Acceptable ranges of model parameters for one \ion{Mg}{2}
and three \ion{C}{4} components (the maximal \ion{C}{4} model) are
summarized in Table~\ref{tab5}.  An example of the best fit model is
overplotted on the observed spectrum in Figure~\ref{fig2}. Here, we
emphasize that the \ion{Mg}{2} component has a relatively high
ionization parameter similar to those of the \ion{C}{4} components.

We also consider a minimal \ion{C}{4} model, for which we need four
more components to reproduce \ion{C}{4} and \ion{Si}{4} because there
is no contribution to the \ion{C}{4} absorption from the \ion{Mg}{2}
phase. We fit the \ion{C}{4} profile with four high-ionization phase
clouds at \delv\ of $-$47, $-$31, $-$12, and 3~\kms\ from the system
center. As with the maximal \ion{C}{4} model, the metallicity of the
\delv\ = $-$47~\kms\ component is constrained to be $\log Z$ $=$
$-$1.0 -- $-$0.9.  We found a range of acceptable ionization
parameters for the first three \ion{C}{4} components, with $\log U$
$=$ $-$1.9 -- $-$1.7.  However, there are no acceptable solutions for
the fourth component (that overlaps with the \ion{Mg}{2}
component). The best model, with $\log Z$ $=$ $-$0.5 and $\log U$ $=$
$-$1.9, reproduces observations for most all transitions, but it gives
rise to additional \ion{Mg}{2} absorption.  These double contributions
from low- and high-ionization phase clouds overproduce the observed
\ion{Mg}{2} profile. This result suggests that the \ion{Mg}{2}
absorption is partially produced by the same cloud as the \ion{C}{4},
and supports the maximal \ion{C}{4} model above. Thus, we conclude
that the maximal \ion{C}{4} model with three \ion{C}{4} components is
more appropriate to describe this system.

As with System~1, the right wing of the \lya\ profile is not produced
either by the \ion{Mg}{2} cloud or by the three \ion{C}{4} clouds.
Therefore, we placed an additional \lya\ line in this region of the
profile.  The unexplained absorption cannot be \lyb\ because this
region is not in the \lyb\ forest, as was the case with System~1.  We
fit the region using one additional \lya\ component with \lognhi\
$\sim$ 14.6 and $b$ $\sim$ 34~\kms. This component is not
well-constrained, but for $\log U$ $>$ $-$2.0 the \ion{C}{4} and
\ion{N}{5} are over-produced unless $\log Z$ $<$ $-$2.0.  For $\log U$
$<$ $-$2.5, the metallicity can be as large as $\log Z$ $=$ 0.0 before
low ionization transitions are over-produced.

We conclude that the $z$ $=$ 1.68079 system in the HE~0429-4091
spectrum has three \ion{C}{4} components as well as one \ion{Mg}{2}
component.  The transitions that provided constraints on $\log U$ and
$\log Z$ based on our Cloudy models are listed in Table~\ref{tab4}.
All components have similar ionization parameters, $\log U$ $\sim$
$-$1.9.  It is difficult to constrain the metallicity of the low
ionization phase for this system since there are kinematically
distributed \ion{C}{4} components that contribute to the \lya\
absorption, however metal line absorption is best reproduced with
about solar metallicity for the \ion{Mg}{2} component.  The blueward,
offset \ion{C}{4} cloud is constrained by the \lya\ profile to have
$\log Z$ $>$ $-$1.1.  Because of its higher ionization state, the size
of the \ion{Mg}{2} cloud of this system ($\sim$ 100~pc) is two orders
of magnitude larger than that of the \ion{Mg}{2} cloud in System~1.
All components are optically thin. Ranges of acceptable model
parameters are listed in Table~\ref{tab5}. An example of the best
models and its model parameters are presented in Figure~\ref{fig2} and
Table~\ref{tab6}. For the model curves shown in Figure~\ref{fig2}, the
nitrogen abundance is decreased by 0.5 dex from the solar abundance.

\subsection{System~3 (HE~2243$-$6031; $z$=1.75570)}
\label{sec:sys3}

\begin{figure}
 \begin{center}
  \includegraphics[width=13cm,angle=0]{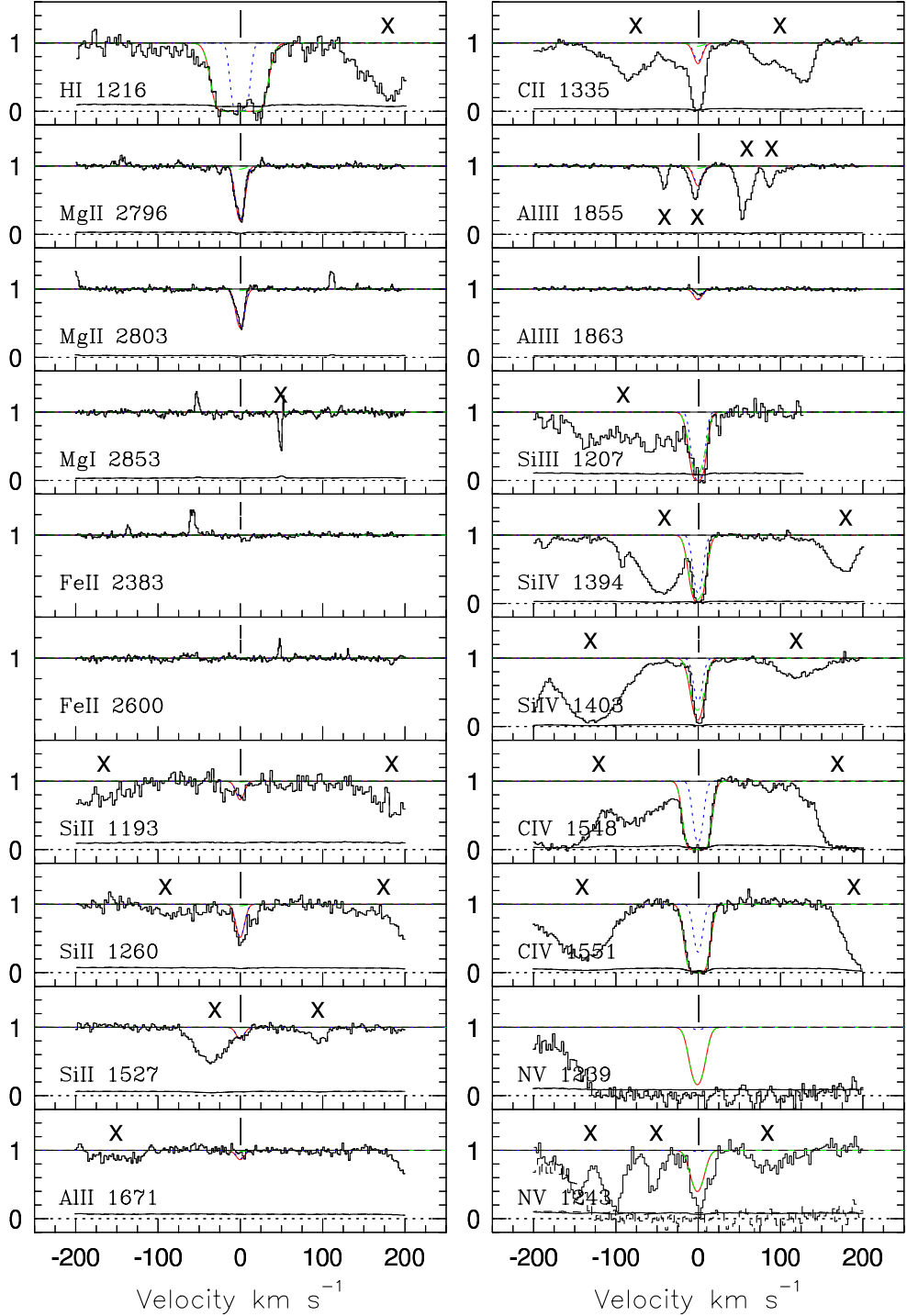}
 \end{center}
 \caption{Same as Figure~\ref{fig1}, but for the system at $z$ = 1.75570
          toward HE~2243$-$6031.\label{fig3}}
\end{figure}

\begin{figure}
 \begin{center}
  \includegraphics[width=15cm,angle=0]{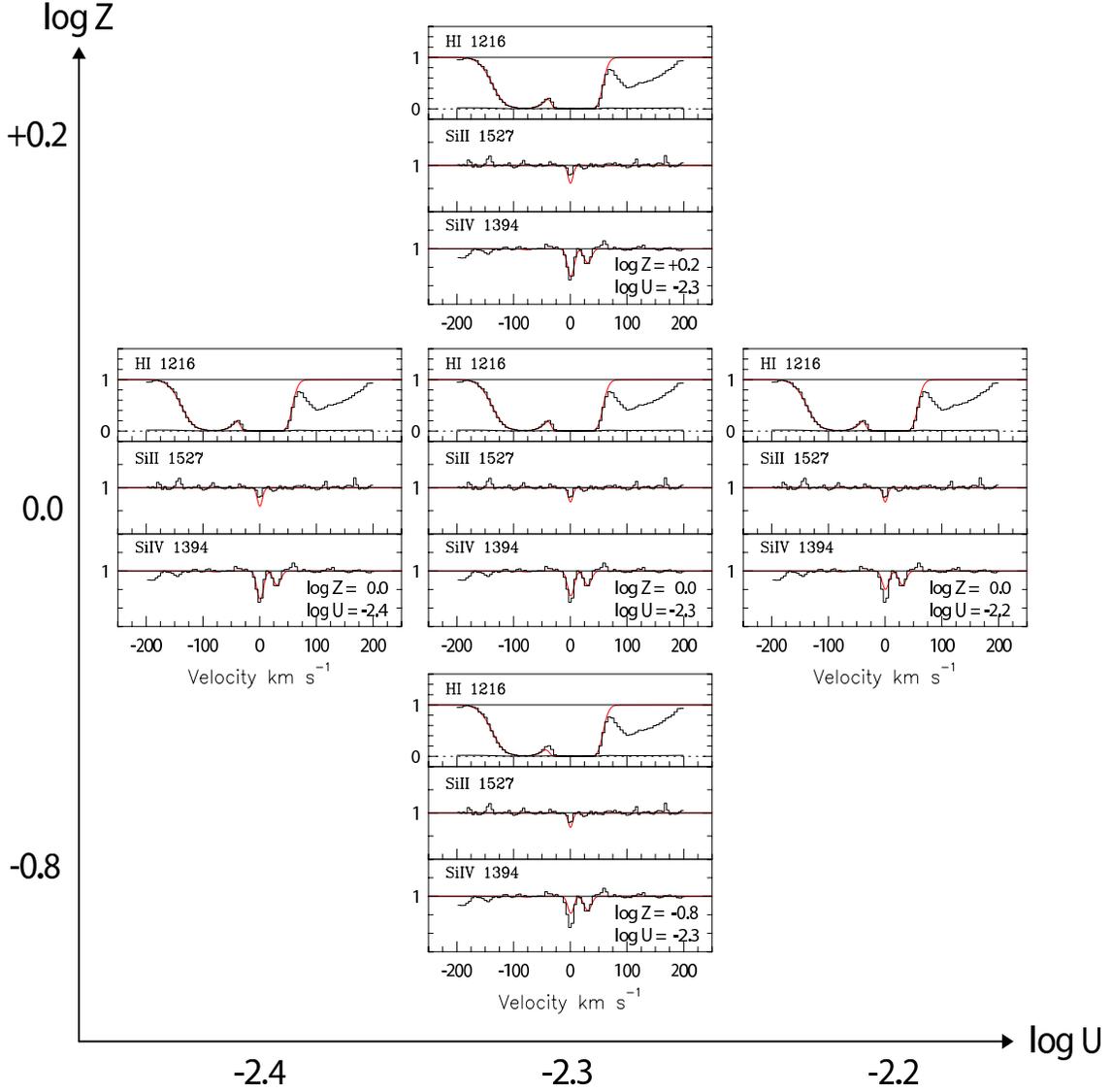}
 \end{center}
 \caption{Synthetic model profiles from Cloudy models superimposed on
          the data for five sets of values of the ionization parameter
          ($\log U$) and metallicity ($\log Z$). The example plotted
          is the $v = 1$~\kms\ cloud of the $z=1.78169$ system toward
          HE~0141-3932.  For other adjacent clouds the model
          parameters set at their preferred values and were not
          adjusted.  The three transitions displayed were those that
          provided the strongest constraints on the parameters for
          this cloud.  The normalized flux scale for the \ion{H}{1}
          panel extends from 0 to 1, but the flux scales for the
          \ion{Si}{2} and \ion{Si}{4} are expanded vertically to
          better show deviations between the models and the data.
          \label{fig4}}
\end{figure}

This weak \ion{Mg}{2} system has simple, single component profiles
detected in \lya, the low-ionization lines, \ion{Mg}{2} and
\ion{Si}{2}, and in the higher-ionization lines, \ion{Al}{3},
\ion{Si}{3}, \ion{Si}{4} and \ion{C}{4}. \ion{Fe}{2} and \ion{Al}{2}
lines are not detected to a 5$\sigma$ equivalent width limit.  See
Table~\ref{tab2} for the rest-frame equivalent widths and $5\sigma$
limit values.  The regions of spectrum where \ion{O}{1}, \ion{C}{2},
and \ion{N}{5} would appear are contaminated by blends, so that they
only serve to provide weak upper limits.  It is immediately apparent
that the equivalent width of \ion{Mg}{2} is very large for this system
compared to its relatively weak \lya\ profile. This implies that the
system has a high metallicity.  Velocity plots of various transitions
are displayed in Figure~\ref{fig3}.

As usual, we first optimize on the column density and Doppler
parameter of \ion{Mg}{2} obtained from Voigt profile fitting (see
Table~\ref{tab3}). A lower limit on the ionization parameter of $\log
U$ $>$ $-$3.5 can be determined, since \ion{Fe}{2} is over-produced at
smaller values.  On the other hand, higher ionization conditions with
$\log U$ $>$ $-$2.0 are ruled out because of overproduction of
\ion{Si}{4}.  The observed \ion{C}{4} absorption is also over-produced
if $\log U$ $>$ $-$1.5.

More stringent constraints on the ionization parameter can be found if
we consider the observed absorption in \ion{Al}{2} and \ion{Si}{2}.
For metallicities $\log Z$ $\lsim$ 0.0, the constraints on ionization
parameter are independent of metallicity.  For larger metallicities,
cooling leads to different constraints.  We will first consider the
constraints for $\log Z$ $<$ 0.0.  Considering \ion{Al}{2} and
\ion{Si}{2}, we found that an ionization parameter at the higher end
of the range $-$3.5 $<$ $\log U$ $<$ $-$2.0 produces the best
agreement, but these ions are still overproduced by the model.  To
reconcile a $\log U$ $=$ $-$2.0 model with the data for $\log Z$ $<$
0.0, we can reduce the abundance of aluminum by 0.5 dex, and that of
silicon by 0.2 dex.  Further variations in abundance pattern could
lead to different preferred values of $\log U$.

Independent of the constraint on the ionization parameter, the \lya\
profile can be used to place a strong lower limit on the metallicity
of the \ion{Mg}{2} cloud.  The \ion{Mg}{2} component is centered
exactly on the \lya\ profile, so that both the blue and the red side
of the \lya\ profile provide the same lower limit, $\log Z$ $>$ 0.0,
in order that \lya\ is not overproduced. 
However, there are still some uncertainties on this metallicity
constraint. On one hand, lower values would be permitted if the system
is in $\alpha$-enhanced condition because this constraint does assume
a solar abundance pattern. On the other hand, the required metallicity
would be higher if magnesium is depleted onto dust grains, although we
do not expect a large amount of dust in weak \ion{Mg}{2} systems as
discussed below.
A cloud with $\log U$ $=$ $-$2.0 and $\log Z$ $=$ 0.0 has a
temperature of $T$ $=$ 9870~K and a size of $1.5$~kpc.  A smaller
$\log U$, which would be possible with further reductions in the
abundances of aluminum and silicon, would lead to a smaller cloud
size.

Metallicities greater than the solar value are also possible for the
\ion{Mg}{2} cloud.  In fact, a solar abundance pattern can be
consistent with the data because the cooling at high metallicity leads
to less \ion{Al}{3} and \ion{Si}{4} production.  If we accept the
constraint on ionization parameter of $\log U$ $\sim$ $-$2.0, as
discussed above, and assume a solar abundance pattern, we find that
\ion{Si}{4} will be slightly over-produced unless $\log Z$ $>$ 0.4.
It is over-produced on the blue side of the \ion{Si}{4}~$\lambda$1403
profile.  For $\log Z$ $>$ 0.4, \ion{Si}{4} is under-produced, but in
this case \ion{Si}{4} absorption can arise from a second, higher
ionization phase.  Similarly, \ion{Al}{3} is over-produced unless
$\log Z$ $>$ 0.9.  (The $\lambda$1863 transition is overproduced; the
$\lambda$1855 transition is contaminated by a blend.)  Due to cooling
at extreme metallicities, a $\log Z$ $=$ 1.0 cloud would have a
temperature of $\sim$ 500~K and a size of $\sim$25~pc.  These
constraints are quite dependent on abundance pattern.  As we noted
above, if the abundance of aluminum was decreased by 0.5 dex relative
to the solar pattern and the silicon abundance was decreased by 0.2
dex, $\log Z$ $=$ 0.0 would be consistent with the data for $\log U$
$=$ $-$2.0.  Both super-solar metallicity models and small changes in
abundance patterns are reasonable ways to reconcile models with the
data.  However, we emphasize that the metallicity constraint of $\log
Z$ $>$ 0.0 is directly determined based on comparison with the \lya\
profile.

Next, we added a high-ionization phase because the broad line profile
of \ion{C}{4} cannot be reproduced by the low-ionization \ion{Mg}{2}
component (see the dotted line in Figure~\ref{fig4} on the \ion{C}{4}
profile).  \ion{Si}{4} is also under-produced by an \ion{Mg}{2}
component with $\log U$ $=$ $-$2.0 and $\log Z$ $=$ 1.0.  In order to
find constraints on the \ion{C}{4} phase, we optimized on the
\ion{C}{4} column density from our Voigt profile fit. In this case,
the contribution to the \ion{C}{4} absorption from the low-ionization
phase was minimal, so that removing it made no significant difference
to the fit, listed in Table~\ref{tab3}.  From Cloudy photoionization
models, we derive a constraint of $\log U$ $=$ $-$1.5 -- $-$1.4 on the
ionization parameter, in order to simultaneously fit the \ion{C}{4}
and \ion{Si}{4} profiles.  If the metallicity of the \ion{Mg}{2} cloud
is lower so that it gives rise to more \ion{Si}{4} absorption, then
the ionization parameter of the \ion{C}{4} cloud would need to be
higher to compensate.  Similarly, if the ionization parameter of the
\ion{Mg}{2} cloud is lower, that of the \ion{C}{4} cloud would also
need to be lower in order to produce more \ion{Si}{4} absorption.

Since the \lya\ profile does not have a contribution from
kinematically separate \ion{C}{4} components, it can be used to place
a lower limit on the metallicity in the high-ionization phase.  In
order that \lya\ is not over-produced by the \ion{C}{4} cloud, a
metallicity of $\log Z$ $>$ $-$0.2 is required.  No firm upper limit
can be placed since the \lya\ absorption would be fully accounted for
by the \ion{Mg}{2} cloud if $\log Z$ $=$ 0.0.  However, if the
\ion{Mg}{2} cloud has $\log Z$ $=$ 1.0 in order to produce the
observed \ion{Al}{3} absorption, it does not account for all of the
observed \lya\ absorption.  In this case, $\log Z$ $=$ $-$0.1 -- 0.0
for the \ion{C}{4} cloud would account for the majority of the \lya\
absorption.  A cloud with $\log U$ $=$ $-$ 1.4 and $\log Z$ $=$ 0.0
has a size of $\sim$ 5~kpc.

We conclude that the system can be reproduced with only two
components: one low-ionization and one high-ionization phase.  The
constraints on $\log U$ and $\log Z$ and the transitions they were
derived from are presented in Table~\ref{tab4} for two models.  Model
1 assumes a solar abundance pattern and relies on a super-solar
metallicity for the low-ionization phase to fit the data.  Model 2
allows an adjustment of the abundances of aluminum and silicon
relative to the solar pattern, but still requires at least a solar
metallicity.  Our favored two-phase model has a relatively high
ionization parameter of $\log U$ $=$ $-$2.0 for the \ion{Mg}{2} cloud,
and $\log U$ $=$ $-$1.4 for the \ion{C}{4} cloud.  However, with a
different abundance pattern, the ionization parameter for the
\ion{Mg}{2} cloud could be lower.  All components of acceptable models
are optically thin in \hi.  Ranges of the acceptable model parameters
are listed in Table~\ref{tab5}, including both models with $\log Z$
$=$ 0.0 and with $\log Z$ $>$ 0.9 for the \ion{Mg}{2} cloud. An
example of best fit models and their model parameters are presented in
Figure~\ref{fig3} (only for the model with $\log Z$ $>$ 0.9) and
Table~\ref{tab6} (for both).

\subsection{Effect of Photoionization Modeling Assumptions}
\label{sec:assumptions}

In using the Cloudy code to derive constraints for the three systems,
we have assumed that photoionization equilibrium applies.  This is
almost certain to be the case for hydrogen since the photoionization
timescale is short, on the order of $10^4$~years.  For metals,
photoionization equilibrium might not apply if the gas has cooled from
a higher temperature, in which case the gas would still be more
ionized than our calculations would suggest.  This would clearly
affect our model constraints.  However, our most important conclusion
of high metallicities for weak \ion{Mg}{2} absorber would only be
stronger if a larger fraction of the magnesium was in higher
ionization states.

Another important assumption we applied in our modeling was that of a
solar abundance pattern, unless a deviation was required by the data.
We note that deviations from the solar pattern are to be expected, and
may apply whether the data require them or not.  In other words, there
is a degeneracy between the parameters $\log U$ and $\log Z$ and the
abundance pattern.
One common type of abundance pattern deviation would be an
$\alpha$-enhanced condition. In this case, we tend to over-estimate
metallicity because our metallicity estimations are based on
magnesium, an $\alpha$-element. Another source of the deviation is
depletion onto dust grains.  We do not really expect a large amount of
dust in weak \ion{Mg}{2} absorbers because they are not typically
close to sites with large \nhi\ or with current star formation.
However, it is important to note that if magnesium is depleted by some
factor, our inferred metallicity would {\it increase} by roughly that
same factor.  

Finally, we consider the effect of changing the shape of the ionizing
radiation field.  For the previous calculations we had assumed a
Haardt and Madau model for the ionizing radiation from quasars and
star forming galaxies, with an escape fraction of 0.1 of ionizing
photons from galaxies \citep{haa96,haa01}.  We have also considered
the opposite extreme, in which only quasars contribute to the ionizing
radiation field.  For a given ionization parameter, the absence of a
stellar contribution leads to an increase in the relative number of
high energy photons.  With that change of spectral shape, we found
negligible difference in the properties derived for the low ionization
phase.  For the high ionization phase, we found that a lower
ionization parameter (e.g., by about 0.5~dex) would be needed to fit
the data.
During the photoionization modeling presented above, we neglected
possible contributions from nearby stellar sources. However, we see
only small differences in the column densities of
low/intermediate-ionization phase gas (e.g., \ion{Si}{2}, \ion{Si}{3},
and \ion{Si}{4}) by a factor of 2 or 3, even after adding stellar
radiation from an {\rm O7} star with an effective temperature of
$T_{eff}$ = 38000~K (model\# C3 of \citealt{sch97}).  We adopt an
ionizing photon number density 10 times greater than that of the
extragalactic background radiation, as a maximum flux model. Thus,
contributions from nearby stellar sources, even if they exist, have no
significant qualitative effect on our conclusions.

We conclude that none of the assumption behind our photoionization
models have a qualitative impact on our conclusions.  In particular,
there is no effect that works against our conclusion of a very high
metallicity for System 3.

\section{Discussion}
\label{sec:discussion}

In this study, we applied photoionization models to constrain the
physical properties of three single-cloud weak \ion{Mg}{2} absorption
systems at $z$ $\sim$ 1.7.  Along with results presented in
\citet{lyn07}, these complete a sample of the seven $z$ $>$ 1.5
single-cloud weak \ion{Mg}{2} absorbers found in the VLT archive
\citep{nar07} for which metallicity constraints could be derived
because of simultaneous coverage of various metal lines and \lya.

We start this final section by summarizing the model constraints
derived for the three $z$ $\sim$ 1.7 single-cloud weak \ion{Mg}{2}
absorbers studied in this paper (\S~\ref{sec:summary}). We then
proceed to compare our results to the other four single-cloud weak
\ion{Mg}{2} systems at $z$ $\sim$ 1.7 from \citet{lyn07} in
\S~\ref{sec:comp_weakMgII}. In this section, we also compare to lower
redshift single-cloud weak \ion{Mg}{2} systems from the literature to
investigate evolutional trends. We address the high metallicity of
these systems in the context of DLAs (whose metallicities are usually
smaller than the solar value by a factor of 10 -- 30) in
\S~\ref{sec:comp_DLA}, sub-DLAs in \S~\ref{sec:comp_subDLA}, and
strong and multiple-cloud weak \ion{Mg}{2} systems (the strong ones
are usually associated with bright ($L$ $>$ 0.05$L^{*}$) galaxies
within 40$h^{-1}$ kpc; e.g., \citealt{ber91}) in
\S~\ref{sec:comp_strongMgII}, respectively.  We finally discuss a
possible correlation between metallicity and total hydrogen column
density in \S~\ref{sec:metal}.

\subsection{Summary of Results}
\label{sec:summary}

System~1 can be fit with a single low-ionization \ion{Mg}{2} cloud and
three higher ionization clouds (two of them offset in velocity from
the \ion{Mg}{2}) that give rise to \ion{C}{4} absorption.  The low
ionization cloud has $\log Z$ $>$ $-$0.8 and two of the high
ionization clouds must have metallicities at least this high as well.
Two offset low-metallicity clouds, producing only \lya\ absorption
appear to be clustered with this system.

System~2 is also fit with one \ion{Mg}{2} and three \ion{C}{4} clouds,
though all four clouds have similar, relatively high ionization
parameters ($\log U$ $\sim$ $-$1.9).  The metallicity of the
\ion{Mg}{2} cannot be well-constrained directly from comparison to the
\lya\ because the kinematical spread of the \ion{C}{4} lines give rise
to a large \lya\ equivalent width, however near solar metallicities
are possible, and even favored for a solar abundance pattern.  Also,
the blueward \ion{C}{4} cloud will over-produce absorption in \lya\
unless its metallicity is $\log Z$ $>$ $-$1.1.  A separate, offset
low-metallicity component was again needed to fit the red portion of
the \lya\ profile.

System~3 has the best metallicity constraint among the seven
single-cloud weak \ion{Mg}{2} absorbers at $z$ $>$ 1.5 because of the
absence of offset \ion{C}{4} components, thereby keeping the \lya\
profile very narrow.  Two phases are still required, with clouds of
different ionization parameters centered at the same velocity that
produce both the narrower low-ionization transitions, and the broader
high-ionization transitions.  Both the low- and the high-ionization
phases are constrained to have solar or super-solar metallicities.

The most striking thing about these model results, is that in all
three $z$ $\sim$ 1.7 systems, the metallicity of at least one phase of
gas is constrained to be greater than one tenth the solar value.  The
three systems all have at least two phases of gas: the low ionization
phase that arises in a layer of gas $\sim$1 -- 100~pc thick with a
density of 0.001 -- 0.1~\cmmm, and the high ionization phase that
comes from a larger region ($\sim$ 0.1 -- 10~kpc) with a lower
density.  For these systems, even metallicities of the larger
high-ionization phase regions range from one tenth solar up to solar.
We will return to a discussion of the surprising issue of how regions
with \lognhi\ $<$ 15 (i.e., over a million times less than the
threshold for star formation) can be enriched in metals to the solar
value.

\subsection{Comparison to Other Single-Cloud Weak \ion{Mg}{2} Systems}
\label{sec:comp_weakMgII}

\begin{figure}
 \begin{center}
  \includegraphics[width=10cm,angle=0]{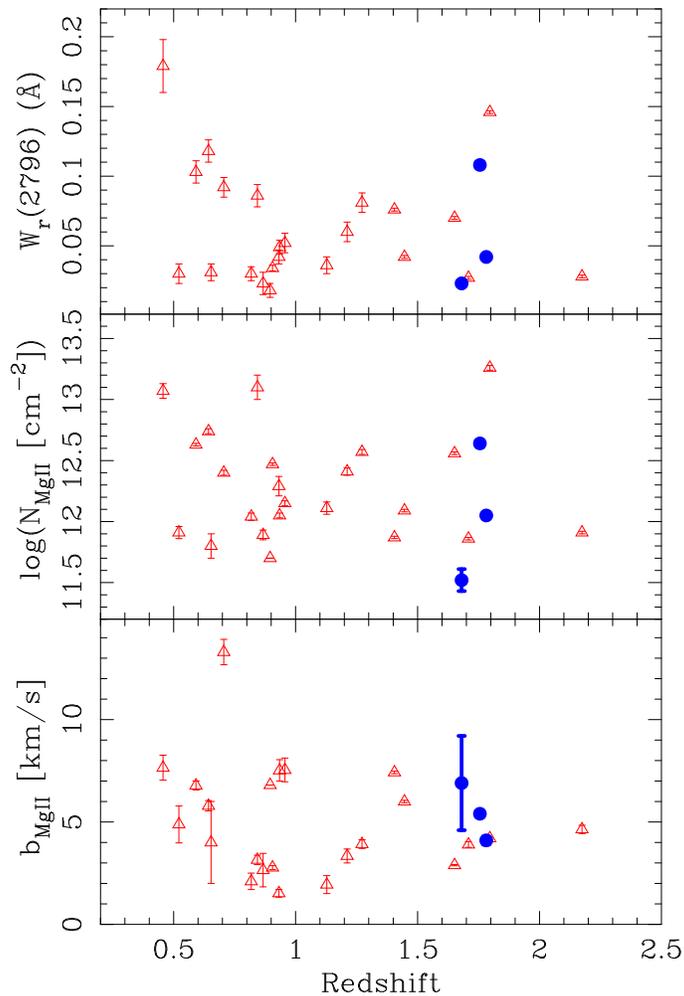}
 \end{center}
 \caption{Rest-frame equivalent width ($W_r$(2796); top panel), Column
          density (\lognmgii; middle panel), and Doppler parameter
          ($b$; bottom panel) of weak \ion{Mg}{2} systems plotted as a
          function of redshift between $z$ = 0.2 -- 2.5. Red points
          (open triangles) and error bars are taken from the
          literature \citep{rig02,cha03,din05,lyn07}, while blue
          points (filled circles) and error bars are from this
          paper.\label{fig5}}
\end{figure}

\begin{figure}
 \begin{center}
  \includegraphics[width=10cm,angle=0]{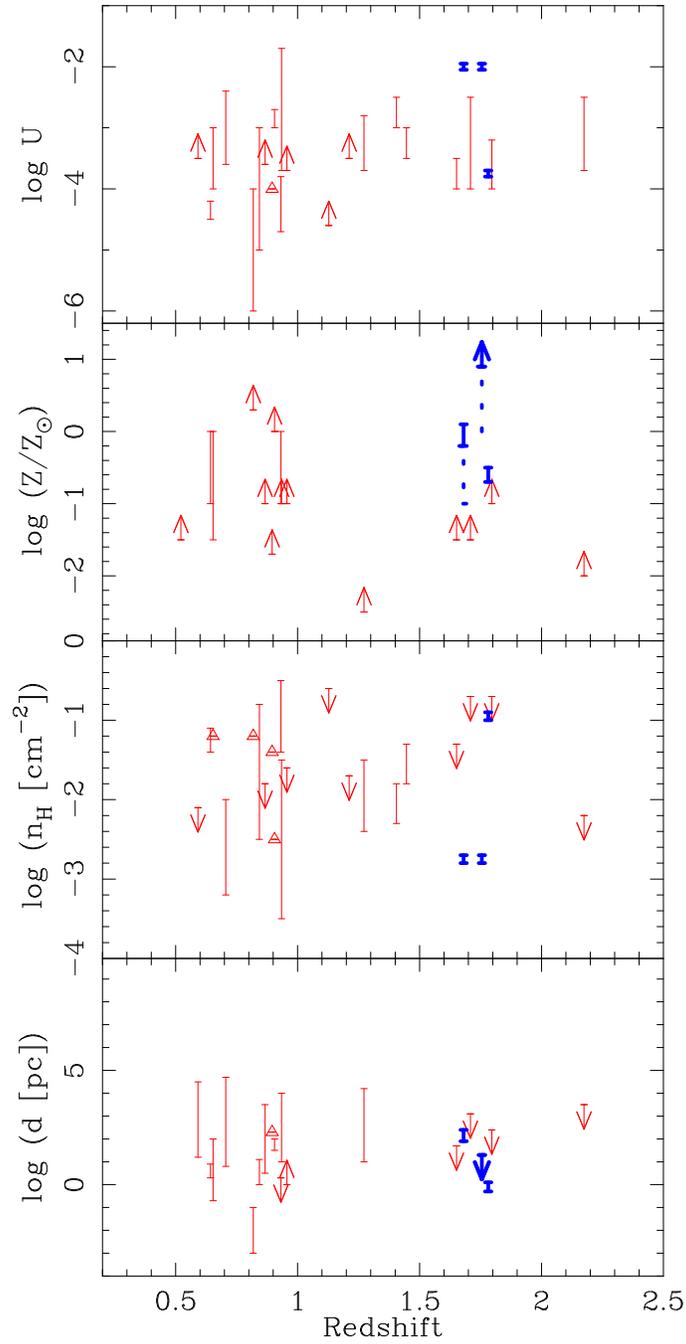}
 \end{center}
 \caption{Same as Figure~\ref{fig5}, but for ionization parameter ($\log
          U$; top panel), metallicity ($\log Z$; 2nd panel), hydrogen
          volume number density ($\log n_{H}$; 3rd panel), and size of
          absorber ($\log d$; bottom panel).\label{fig6}}
\end{figure}

The properties of the three $z$ $\sim$ 1.7 single-cloud weak
\ion{Mg}{2} systems, derived from the observed profiles, are compared
to those of 23 other single-cloud weak \ion{Mg}{2} absorbers in
Figure~\ref{fig5}, both at similar and at lower redshifts.
Figure~\ref{fig5} shows that there is no significant evolution in the
observed \ion{Mg}{2}~$\lambda$2796 profiles.  Figure~\ref{fig6}
presents quantities derived from photoionization models, for our three
absorbers and for others taken from the literature, using the same
methods as we have used here \citep{rig02,cha03,din05,lyn07}.  The
specific constraints that are displayed in Figures~\ref{fig5} and
\ref{fig6} are also listed in Table~\ref{tab7}, with the relevant
references.

\citet{lyn07} derived similar constraints for the phase structure of
four different single-cloud weak \ion{Mg}{2} absorbers at $z$ $\sim$
1.7.  They also placed constraints on the metallicities of those
absorbers, of $\log Z$ $>$ $-$1.5, $>$ $-$1.5, $>$ $-$1.0, and $>$
$-$2.0 for the four different low-ionization phases.  They note that
the metallicities could be significantly higher than these lower
limits, however, these absorbers (by chance) tended to have larger
contributions to the \lya\ absorption from offset \ion{C}{4} clouds
and from \lya-only clouds.

Results from modeling lower redshift (0.4 $<$ $z$ $<$ 1.4)
single-cloud weak \ion{Mg}{2} systems also typically yield a two-phase
structure, with similar densities.  Metallicity constraints are again
often limited because of the difficulty in separating the
contributions of the two phases.  However, of the 11 cases of
single-cloud weak \ion{Mg}{2} absorbers for which some metallicity
constraint could be obtained, 2 cases require the metallicity of the
low-ionization phase to be be solar or even super-solar \citep{cha03}.
Furthermore, a total of 7/11 of the low redshift cases required a
metallicity of at least one tenth of the solar value.  We note that
the metallicities are often quite likely to be higher than these
strict lower limits because they are derived assuming that none of the
\lya\ absorption comes from the high-ionization clouds.

We conclude that some fraction of weak \ion{Mg}{2} systems have been
demonstrated to have solar or super-solar metallicity, at low redshift
(2 systems at \zabs\ = 0.8181 and 0.9056; \citealt{cha03}) and even at
$z$ $\sim$ 1.7 (1 system at \zabs\ = 1.7557; this paper), which
corresponds to an age of the Universe of 3.7~Gyr.  Values constrained
to be greater than 1/10th the solar value are even more common.
Furthermore, it is likely that in many cases for which lower limits on
metallicity have been derived the actual value is much higher.  The
metallicity constraint that we have derived for the \zabs\ = 1.7557
absorber in this paper is high because this system is a rare case that
does not have offset \ion{C}{4} absorption that contributes to the
\lya\ absorption.  In other ways there is no reason to think it is
unusual.  The same applied for the low redshift weak \ion{Mg}{2}
systems with the highest metallicity constraints \citep{cha03}.  We
therefore know that single-cloud weak \ion{Mg}{2} absorbers at both
low and high redshift definitely have metallicities of at least 1/10th
solar, but that some and probably many have solar or even supersolar
metallicities.

Thus even before $z=1.5$ star formation must have polluted certain
environments with metal-rich gas.  Since \citet{lyn07} could not
constrain any $z$ $>$ 1.5 single-cloud weak \ion{Mg}{2} absorbers to
have close to solar metallicity, and since several lower redshift
solar metallicity cases were known, they tentatively suggested that
there might be a gradual build-up of metals in the single-cloud weak
\ion{Mg}{2} absorber population.  In this paper, we have shown there
is at least one strong counter-example, i.e. of $z$ $>$ 1.5 solar
metallicity cases, and therefore there appears to be no metallicity
evolution in the population.  This is not to say that the population
has a narrow range of physical properties: there is a large spread of
cloud densities and metallicities at all redshifts (see
Figure~\ref{fig6}).  However, it does suggest that common processes
are at work to create the metals, and that the metals are mixed into
fairly similar surrounding environments.

The existence of high metallicity (greater than solar) compact
intergalactic clouds ($\sim 100$~pc) has also been demonstrated by
\citet{sch07}, who derived robust lower limits on metallicities of
\ion{C}{4} absorbers using similar photoionization modeling
techniques.  \citet{sch07} find that the number density of this
population of high metallicity \ion{C}{4} absorbers is of the same
order as that of single-cloud weak \ion{Mg}{2} absorbers, and suggests
that the former might evolve from the latter as material expands.
More observations allowing direct connections between these
populations will be of great interest.  In fact, two of our three $z$
$>$ 1.5 weak \ion{Mg}{2} systems have \ion{C}{4} phases with
metallicities comparable to those derived by \citet{sch07} for the
high metallicity \ion{C}{4} cloud population.

\subsection{Comparison to Damped \lya\ Systems}
\label{sec:comp_DLA}

Damped \lya\ systems (DLAs), whose neutral hydrogen column densities
are greater than \lognhi\ $=$ 20.3, dominate the neutral gas in the
Universe at high redshift \citep[e.g.,][]{lan95,sto00,per03}. By
identifying metal absorption lines that correspond to high redshift
DLAs, it is possible to constrain the global metal abundances and
trace the evolution of metallicity in neutral gas from $z$ $=$ 5 to 0
\citep[e.g.,][]{pet94,pro00}.

The metallicity evolution, measured from DLAs, however, has two big
problems: (i) the global mean metallicity is less than 1/6th of the
solar abundance at $z$ $>$ 1 \citep[e.g.,][]{pet99,pro00}, and is
still less than the solar by a factor of $>$ 4 if the metallicity
redshift relationship is extrapolated down to $z$ $\sim$ 0 using a
least-squares linear fit \citep{pro03,kul05,per06b}
\footnote{At $z$ $<$ 1, there are only a few cases of metallicity
measurements in DLAs using high quality spectra, and in these cases
the metallicity is evaluated to be $\sim$ 1/10th of the solar value
\citep{var00,pet00}}, and (ii) the metallicity is also much smaller
than the theoretically expected value from the star forming activity
at higher redshift \citep[e.g.,][]{mad98,pet99,wol03}.  This
discrepancy (i.e., ``missing-metals problem'') remains an unresolved
problem.

There are at least two possible classes of ideas proposed for the
origin of this discrepancy: (i) a large amount of dust in metal-rich
DLAs obscures background quasars, which prevent us from detecting DLAs
with higher metallicities \citep[e.g.,][]{boi98,fal93,vla05}, and (ii)
the DLA region is sampling a lower metallicity than is found in the
star-forming parts of the galaxy \citep{ell05a,wol03}.

The dust obscuration scenario is supported by the small abundances of
depleted elements such as Cr and Fe, relative to undepleted elements
like Zn, in DLA systems \citep[e.g.,][]{pet97,kha04}. \citet{vla05}
estimated that $\sim$30~\%\ to 50\%\ DLAs are missed as a consequence
of obscuration.  In absorbers expected to have a lower dust abundance,
sub-DLA systems (super-LLSs) with \lognhi\ $=$ 19.0 -- 20.3, a higher
metallicity has been measured
\citep[e.g.,][]{per06a,pro06,pet00,kul07}. Although \citet{ell01} did
not see any significant differences in DLAs toward radio-selected
quasars, compared to those toward optically selected quasars at $z$
$>$ 2, the dust-depletion effect is expected to be more important at
lower redshift.

Alternatively, \citet{ell05a} and \citet{che05} suggest that there
could be an metallicity gradient as a function of a distance from the
galactic center, with DLAs produced at a larger impact parameter than
the stellar emission.  DLAs that are produced in the host galaxies of
gamma-ray bursts tend to have higher metallicities
\citep[e.g.,][]{djo04}, which could be more representative of typical
star-forming regions.  Also, small regions with DLA column densities
could be ejected from galaxies by superwinds or as a part of AGN
outflows \citep{mac99,ham97,gab06,bou06}.

Figure~\ref{fig7} summarizes the metallicities derived for DLAs at $z$
$=$ 0.4 to 4.8 from \citet{pro03} and \citet{kul05}, and compares to
the values for single-cloud weak \ion{Mg}{2} absorbers.  It is
striking that weak \ion{Mg}{2} absorbers, despite having drastically
smaller \nhi\ values, have considerably higher metallicities than the
DLAs, both at low redshift and at $z$ $\sim$ 1.7.

Because they are not near regions with current star formation, the
dust-to-gas ratio in single-cloud weak \ion{Mg}{2} absorbers is
expected to be very small.  Even for DLAs, it has been demonstrated
that there is no bias against observing high metallicities due to dust
obscuration of their background quasars \cite{ell01,ell05b}.  Even if
the dust-to-gas ratio is similar to those in DLAs, their total dust
column density should be much smaller than those in DLAs, because of
their lower gas column densities.  Thus, they would not be subject to
the metallicity bias that was more likely to have affected DLAs.  Thus
the weak \ion{Mg}{2} absorbers provide an opportunity to see some high
metallicity regions of the universe at high redshift, though they are
not the same type of regions that would produce high metallicity DLA
absorption.

Relating to the idea that DLAs sample low metallicity parts of
galaxies, the weak \ion{Mg}{2} absorbers, despite their low column
densities, must somehow sample higher metallicity regions which are
not closely related to luminous galaxies.  This provides clear
evidence for metallicity inhomogeneities, which because of the size
constraints for single-cloud weak \ion{Mg}{2} absorbers, are on very
small scales (parsecs to hundreds of parsecs).

\subsection{Comparison to sub-DLA Absorbers}
\label{sec:comp_subDLA}

Sub-DLA absorbers, with \lognhi\ $=$ 19.0 -- 20.3, have been found to
have systematically higher metallicities than DLAs with the
\nhi-weighted value larger by 0.5 -- 0.8 dex, at 0.6 $<$ $z$ $<$ 3.2
\citep{kul07}.  Several sub-DLA systems (super-LLSs) at 0.5 $<$ $z$
$<$ 2.0 have been found with metallicities that are solar or
super-solar \citep[e.g.,][]{per06a,pro06}.  Even small numbers of such
high metallicity systems contribute significantly to the total metal
mass density at high redshift \citep{pro06,kul07}.  The \citet{kul07}
mean values of metallicity for sub-DLAs, determined from [Zn/H], are
reproduced in our Figure~\ref{fig7} from their Table~1.  Although
there is a range of metallicities at each redshift, there is a clear
increase in the metallicity of sub-DLAs from $\langle z \rangle$ $=$
1.5 to $\langle z \rangle$ $=$ 0.9.  This is consistent with the
global star formation history in the universe over this period, and
the metallicity values are much more in line with expectations from
chemical evolution models than those for DLAs \citep{kul07,som01}.

Some single-cloud weak \ion{Mg}{2} absorbers, with \lognhi\ $\sim$ 15
-- 16, also have solar or super-solar metallicities, both at $\langle
z \rangle$ $=$ 0.9 and at $\langle z \rangle$ $=$ 1.5.  There is
considerable overlap between the metallicity constraints for sub-DLAs
and single-cloud weak \ion{Mg}{2} absorbers at low and intermediate
redshifts.  However, it is important to note that many of the weak
\ion{Mg}{2} metallicities are only lower limits, and it is quite
likely that the values are higher at least in some cases.  The reason
that we cannot derive higher metallicities for those systems could
simply be the separate \ion{C}{4} clouds that happen to overlap in
velocity.  
It is also important to note that for $z$ $\sim$ $1.7$, the redshift
of the absorbers we have studied in this paper, there are not many
sub-DLA systems with super-solar metallicity, but all single-cloud
weak \ion{Mg}{2} absorbers potentially could have solar- or
super-solar metallicity because we can place only lower limits on
metallicity. Although two super-solar metallicity sub-DLAs were
investigated by \citet{pro06}, they were selected for study from a
much larger sample because of their extremely strong \ion{Zn}{2}
absorption. However, much larger data samples of sub-DLAs and weak
\ion{Mg}{2} systems will be required before concluding whether there
are any differences of metallicity between these two categories.

Another apparent difference between the metallicities of single-cloud
weak \ion{Mg}{2} absorbers and sub-DLAs, is that there is no clear
increase in the metallicity of the single-cloud weak \ion{Mg}{2}
absorbers over the range 0.4 $<$ $z$ $<$ 1.7.  This type of absorption
is apparently produced by the same types of regions present over this
redshift range, though they are known to be less common at low
redshifts and at $z$ $>$ 2 \citep{nar05,nar07,lyn06,lyn07}.

The sample size of weak \ion{Mg}{2} systems is now much smaller than
those for DLAs and Sub-DLAs, and only a few cases have been confirmed
to have super-solar metallicities. It is quite important to increase
the sample size of weak \ion{Mg}{2} absorbers, particularly to include
other systems with only one \ion{C}{4} component such that a strong
constraint on metallicity can be derived.  At $z < 1.5$ this will
involve high resolution observations with {\it HST}/COS or {\it
HST}/STIS of quasars for which high-resolution optical spectra are
also available.

\subsection{Comparison to Strong \ion{Mg}{2} and Multiple-Cloud Weak \ion{Mg}{2} Absorbers}
\label{sec:comp_strongMgII}

Intermediate in \lognhi\ between single-cloud weak \ion{Mg}{2}
absorption and sub-DLA's are Lyman limit and partial Lyman limit
systems, with \lognhi\ $=$ 16 -- 19.  These populations loosely
correspond to the bulk of the strong \ion{Mg}{2} absorber population,
with $W_r$(2796) $>$ 0.3~{\AA}, and to multiple-cloud weak \ion{Mg}{2}
absorbers.  Fewer of these systems have been studied in detail to
date, but it is still possible to derive good constraints on
metallicities in many cases, using photoionization modeling techniques
the same as those employed here.  The ten cases of strong and
multiple-cloud weak \ion{Mg}{2} absorbers at 0.7 $<$ $z$ $<$ 1.9
\citep{mas05,zon04,lyn07,pro99,din03}, shown on Figure~\ref{fig7},
have metallicities ranging from $-$1.5 $<$ $\log Z$ $<$ 0.6.  These
are consistent with the ranges for both sub-DLAs and single-cloud weak
\ion{Mg}{2} absorbers, but tend to be higher than those for DLAs.
Without larger samples, we cannot distinguish differences or
evolutionary trends, but again we note that many of these are
measurements of the metallicity, while for the single-cloud weak
\ion{Mg}{2} absorbers the low metallicity values are all lower limits.

\subsection{Metallicity vs. \nhi}
\label{sec:metal}

\begin{figure}
 \begin{center}
  \includegraphics[width=15cm,angle=0]{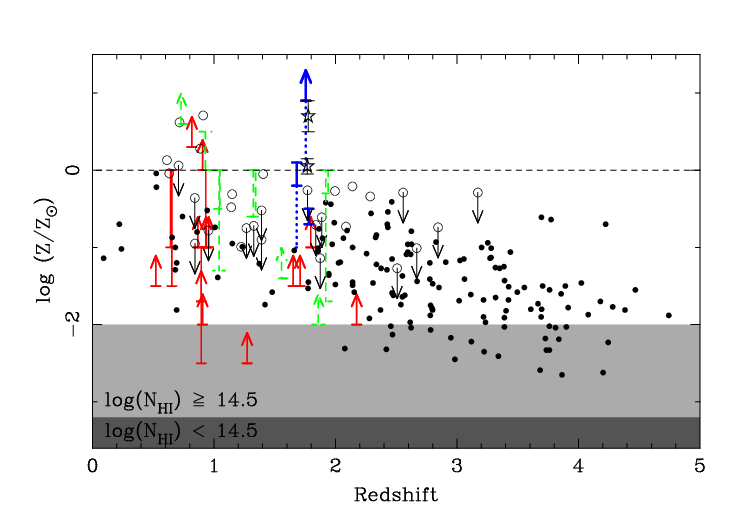}
 \end{center}
 \caption{Metallicity in single-cloud weak \ion{Mg}{2} systems as a
          function of redshift (marked with solid red/blue error
          bars), compared to that of 148 DLA systems (black points;
          \citealt{pro03} and \citealt{kul05}), 30 sub-DLA systems
          (open circles; \citealt{kul07}), 10 Lyman limit systems
          including strong \ion{Mg}{2} and multiple-cloud weak
          \ion{Mg}{2} systems (dashed green error bars;
          \citealt{lyn07,mas05,zon04,din03,pro99}), Lyman $\alpha$
          forest clouds with \lognhi\ $\geq$ 14.5 (thin shaded region;
          \citealt{son96,cow95,tyt95}), and Lyman $\alpha$ forest
          clouds with \lognhi\ $<$ 14.5 (thick shaded region;
          \citealt{cow98,lu98}).  At $z$ $\sim$ 1.7, the three thick
          error bars with dotted lines are the more conservative
          metallicity constraints for the single-cloud weak
          \ion{Mg}{2} systems that we have modeled (using \lya\ as a
          direct constraint), and the two open stars are the
          super-solar sub-DLA systems from \citet{pro06}.  The
          horizontal dotted line denotes the solar
          metallicity.\label{fig7}}
\end{figure}

In summary, Figure~\ref{fig7} shows that it is the lowest \nhi\
systems, that appear to have the highest metallicities, both at $z$
$<$ 1 and at $z$ $\sim$ 1.7.  The data are even consistent with a
gradual increase in metallicity with decreasing \nhi, though there is
a huge spread at any given value. Such a trend has already been
pointed out using compiled samples of DLAs and sub-DLAs
\citep{boi98,ake05,mei06,kha07}, extending down to sub-DLAs with
\lognhi\ $\sim$ 19 \citep{per03}. \citet{yor06} also suggested that a
similar trend continued to strong \ion{Mg}{2} systems with smaller
hydrogen column densities.

The individual systems plotted on Figure~\ref{fig7}, however, are only
the systems with detected \ion{Mg}{2} absorption.  In particular, for
single-cloud weak \ion{Mg}{2} absorbers we are sure to be missing
objects with metallicities significantly less than those of the
systems that we do detect.  In fact, those lower metallicity objects
are part of the much larger \lya\ forest population.
Figure~\ref{fig7} also shows the typical metallicities of \lya\ forest
clouds, both those with \lognhi\ $>$ 14.5, which have $\log Z$ \lsim\
$-$2.0 \citep{son96,cow95,tyt95}, and those with lower column
densities, which have a mean metallicity of $\log Z $ \lsim\ $-$3.2
\citep{cow98,lu98}.  These much lower values for the \lya\ forest
overlap with the DLA metallicities at $z$ $>$2.5, but are smaller than
those for low redshift DLAs.

The fact that the high metallicity population of single-cloud weak
\ion{Mg}{2} absorbers represents only a small fraction of all \lya\
forest clouds does not diminish their significance.  The single-cloud
weak \ion{Mg}{2} absorbers at $z$ $\sim$ 1 would account for 25 --
100\%\ of the \lya\ forest clouds with 15.8 $<$ \lognhi\ $<$ 16.8
\citep{rig02}.  More importantly, the cross-section of single-cloud
weak \ion{Mg}{2} absorbers in the plane of the sky is comparable to
that of galaxies at $z$ $\sim$ 1, considering the regions of galaxies
$\lsim$30 -- 40~kpc that produce Lyman limit absorption.  Thus there
are significant regions of the universe covered by these mysterious
near-solar or super-solar metallicity objects with small \lognhi.

As described above, one explanation proposed for the larger
metallicities for sub-DLAs than for DLAs is the dust obscuration of
quasars that have the highest metallicity DLAs in their foregrounds.
Although there may be some debate about whether this is a plausible
explanation of that trend, it is not likely to explain the continued
increase in metallicity toward the lowest \lognhi\ weak \ion{Mg}{2}
absorbers.  There is not likely to be a large selection effect due to
dust obscuration for even the sub-DLA systems.  It seems more
plausible that different types of absorbers are sampling different
types of regions in and around galaxies, which can have large
variations in metallicities even on small scales.  Different types of
galaxies, e.g. star-forming versus quiescent, will have different
fractions of area covered by regions of the different types. By
comparing to the mass-metallicity relation seen in star-forming
galaxies at local universe \citep{tre04} and higher redshift
\citep{sav05,erb06}, \citet{yor06} and \cite{kha07} proposed that the
DLAs are associated with low-mass ($<$ $10^9$ $M_{\odot}$) galaxies,
while sub-DLAs and weaker LLSs are probably the systems that arise in
massive spiral/elliptical galaxies.

The question remains: how do the single-cloud weak \ion{Mg}{2}
absorbers develop such high metallicities even at redshifts as high as
$z$ $\sim$ 1.7?  We know that the lines of sight that pass through
these objects have \nhi\ five or six orders of magnitude below the
star formation threshold.  We also know that most single-cloud weak
\ion{Mg}{2} absorbers are not located very close to luminous galaxies,
but that they tend to be in the vicinity \citep{chu05,mil06}.
Outflowing gas from dwarf galaxies in the local universe
\citep[e.g.,][]{sto04} or from Lyman break galaxies at higher redshift
\citep[e.g.,][]{ade05} could partially contribute to the high
metallicities in single-cloud weak \ion{Mg}{2} systems.  \citet{sch07}
have argued that high metallicity intergalactic \ion{C}{4} clouds may
be ejected as small clumps in superwinds from star-forming galaxies.
However, at least some of the very high metallicity single-cloud weak
\ion{Mg}{2} absorbers show signs of \ion{Fe}{2} line detections
(e.g. our System~1 and several systems in \citealt{rig02}), although
sometimes it is detected with only a few $\sigma$ level. For those
systems, it has been stated that they are not $\alpha$-enhanced, and
that "in situ" star formation, including the less energetic Type~Ia
SNe that increase the iron to magnesium ratio, is responsible for
enriching the gas \citep{rig02}.
Here, readers should be aware that it is not the gas along the weak
\ion{Mg}{2} absorber line of sight that has formed stars, since it has
much too low a column density.  It must instead be a nearby molecular
cloud region that was actively star forming more than 1 billion years
before the time from which we observe the absorber.  That region would
itself produce DLA absorption if a line of sight passed through it.
It is known that many DLAs at $z$ $<$ 1 are associated only with dwarf
or low surface brightness galaxies \citep{rao03}, and that in at least
one case no galaxy is detected in deep narrow-band H$\alpha$ images
\citep{bou01}.

So then why do DLAs not have high metallicities as well?  And how do
we explain the general trend of increasing metallicity with decreasing
\nhi?  The reason could be simple.  Large quantities of metals can be
produced in star forming regions in a variety of environments, inside
and outside of galaxies.  However, the metals will be dispersed into
widely different types of regions, depending upon the environment.
Within giant galaxies molecular cloud regions are typically surrounded
by a significant volume of high density, high column density gas.  So
the metallicity is diminished by a large dilution effect as the
ejected metals are spread through the galaxy region.  In an
intergalactic star-forming region, which is metal-polluted by {\it
in-situ} star formation or has been ejected from a nearby dwarf
galaxy, the clouds may not have a significant volume of high column
density gas around it, thus the dilution could be minimal.  This would
especially apply in a shallow potential well, where much of the
surrounding gas is ejected by the early Type~II SNe, and the remnant
low density gas remaining is greatly enhanced by the metals produced
by later Type~Ia's which encounter a relatively sparse surrounding
medium.  The same type of dilution effect could lead to DLAs having
lower metallicity than sub-DLAs.  \citet{pro06} suggests that the
super-solar metallicity sub-DLAs that they observed at $z$ $\sim$ 1.7,
extreme members of the sub-DLA population, are related to regions
affected by the feedback from actively star-forming regions.  The
metals would be spread to larger lower density sub-DLA regions by
energetic processes.

Obviously, the metallicity of a region is determined by a combination
of the amount of star formation activity in its vicinity and the
overall density of its gas.  In the case of the single-cloud weak
\ion{Mg}{2} absorbers, we conclude that the density effect dominates,
leading to unexpected patches of high metallicity outside of galaxies.
In the contrasting case of a DLA, there is just too much gas around it
to dilute the metals coming through it, and a low metallicity often
results.

\acknowledgments 
This research was funded by the National Science Foundation (NSF)
under grant AST-04-07138 and by NASA under grant NAG5-6399. This work
was also partially supported by the Sumitomo foundation (070380). We
also acknowledge the ESO archive facility for providing the data. This
paper benefit from many constructive suggestions by an anonymous
referee.

\clearpage


\begin{deluxetable}{cccccccccc}
\tabletypesize{\scriptsize}
\tablecaption{Observation Log for Three Quasars \label{tab1}}
\tablewidth{0pt}
\tablehead{
\colhead{QSO}           &
\colhead{RA}            &
\colhead{Dec}           &
\colhead{$z_{em}$}      &
\colhead{$m_{V}$}       &
\colhead{$\lambda$$^a$} &
\colhead{Setting$^b$}   &
\colhead{$t_{exp}$}     &
\colhead{Date}          &
\colhead{Prog. ID$^c$}  \\
\colhead{}              &
\colhead{(h:m:s)}       &
\colhead{(d:m:s)}       &
\colhead{}              &
\colhead{(mag)}         &
\colhead{(\AA)}         &
\colhead{}              &
\colhead{(sec)}         &
\colhead{(yyyy mm dd)}  &
\colhead{}              \\
\colhead{(1)}           &
\colhead{(2)}           &
\colhead{(3)}           &
\colhead{(4)}           &
\colhead{(5)}           &
\colhead{(6)}           &
\colhead{(7)}           &
\colhead{(8)}           &
\colhead{(9)}           &
\colhead{(10)}          
}
\startdata
HE~0141$-$3932 & 01:43:33.70 & $-$39:17:01.0 & 1.807 & 16.3 & 3060 -- 10000 & 437 $\times$ 860 & 14400 & 2001 07 19 -- 2001 08 24 & 67.A-0280 \\ 
               &             &               &       &      &               & 346 $\times$ 580 & 25200 &                          &           \\
HE~0429$-$4901 & 04:30:37.39 & $-$48:55:24.2 & 1.940 & 16.2 & 3050 -- 10080 & 437 $\times$ 860 & 18835 & 2001 01 13 -- 2001 03 19 & 66.A-0221 \\ 
               &             &               &       &      &               & 346 $\times$ 580 & 10800 &                          &           \\
HE~2243$-$6031 & 22:47:08.85 & $-$60:15:46.9 & 3.010 & 18.3 & 3140 -- 10000 & 346 $\times$ 580 & 14400 & 2000 06 04 -- 2000 06 12 & 65.O-0411 \\ 
               &             &               &       &      &               & 437 $\times$ 860 & 11400 &                          &           \\
\enddata 
\tablenotetext{a}{Observed wavelength range.}
\tablenotetext{b}{Central wavelength for blue and red CCD chips of VLT/UVES.}
\tablenotetext{c}{Proposal ID used in the ESO webpage.}
\end{deluxetable}

\begin{deluxetable}{lccc}
\tablecaption{Rest-Frame Equivalent Widths \label{tab2}}
\tablewidth{0pt}
\tablehead{
\colhead{} &
\multicolumn{3}{c}{Equivalent Width, $W_{rest}$ (\AA)$^a$} \\
\cline{2-4} 
\colhead{}           &
\colhead{System~1}   &
\colhead{System~2}   &
\colhead{System~3}   \\
\colhead{Transition} &
\colhead{(\zabs\ = 1.78169)}   &
\colhead{(\zabs\ = 1.68079)}   &
\colhead{(\zabs\ = 1.75570)}   
}
\startdata
 \lya                       &     0.991 $\pm$ 0.003 &     0.816 $\pm$ 0.009 &        0.341 $\pm$ 0.008 \\
 \ion{Mg}{2}~$\lambda$2796  &     0.039 $\pm$ 0.001 &     0.015 $\pm$ 0.001 &        0.121 $\pm$ 0.003 \\
 \ion{Mg}{2}~$\lambda$2803  &     0.020 $\pm$ 0.001 &     0.008 $\pm$ 0.001 &        0.059 $\pm$ 0.001 \\
 \ion{Mg}{1}~$\lambda$2853  & $<$ 0.001             & $<$ 0.001$^b$         &        0.007 $\pm$ 0.001 \\
 \ion{O }{1}~$\lambda$1302  & $<$ 0.001             & $<$ 0.001             &    $<$ 0.147$^b$         \\
 \ion{Fe}{2}~$\lambda$2383  & $<$ 0.001             & $<$ 0.001             &        0.009 $\pm$ 0.002 \\
 \ion{Fe}{2}~$\lambda$2600  &     0.002 $\pm$ 0.001 & $<$ 0.001             &    $<$ 0.001             \\
 \ion{Si}{2}~$\lambda$1193  &     0.018 $\pm$ 0.002 & $<$ 0.003             & $\sim$ 0.015$^b$         \\
 \ion{Si}{2}~$\lambda$1260  &     0.020 $\pm$ 0.001 &     0.008 $\pm$ 0.002 & $\sim$ 0.040$^b$         \\
 \ion{Si}{2}~$\lambda$1527  &     0.004 $\pm$ 0.001 & $<$ 0.001             & $\sim$ 0.019$^b$         \\
 \ion{Al}{2}~$\lambda$1671  &     0.006 $\pm$ 0.001 &     0.003 $\pm$ 0.001 &    $<$ 0.002             \\
 \ion{C }{2}~$\lambda$1335  &     0.028 $\pm$ 0.001 &     0.012 $\pm$ 0.001 &    $<$ 0.438$^b$         \\
 \ion{Al}{3}~$\lambda$1855  &     0.004 $\pm$ 0.000 &     0.005 $\pm$ 0.001 &    $<$ 0.045$^b$         \\
 \ion{Al}{3}~$\lambda$1863  &     0.002 $\pm$ 0.000 & $<$ 0.001             &        0.006 $\pm$ 0.001 \\
 \ion{Si}{3}~$\lambda$1207  & $<$ 0.147$^b$         &     0.114 $\pm$ 0.008 &        0.101 $\pm$ 0.004 \\
 \ion{Si}{4}~$\lambda$1394  &     0.032 $\pm$ 0.002 &     0.110 $\pm$ 0.002 &        0.100 $\pm$ 0.001 \\
 \ion{Si}{4}~$\lambda$1403  &     0.019 $\pm$ 0.002 &     0.069 $\pm$ 0.002 &        0.081 $\pm$ 0.001 \\
 \ion{C }{4}~$\lambda$1548  &     0.082 $\pm$ 0.002 &     0.250 $\pm$ 0.002 &        0.196 $\pm$ 0.002 \\
 \ion{C }{4}~$\lambda$1551  &     0.048 $\pm$ 0.002 &     0.171 $\pm$ 0.002 &        0.174 $\pm$ 0.003 \\
 \ion{N }{5}~$\lambda$1239  & $<$ 0.001             & $<$ 0.002             &             $^c$         \\
\enddata 
\tablenotetext{a}{errors include both photon noise and continuum level
                  uncertainty.}
\tablenotetext{b}{blending with other lines.}
\tablenotetext{c}{flux is completely black.}
\end{deluxetable}
\clearpage

\begin{deluxetable}{ccccc}
\tablecaption{Voigt Profile Fits for Detected Transitions \label{tab3}}
\tablewidth{0pt}
\tablehead{
\colhead{Transition}    &
\colhead{\delv}         &
\colhead{$\log N$}      &
\colhead{$b$}           &
\colhead{optimized$^a$} \\
\colhead{}              &
\colhead{(\kms)}        &
\colhead{}              &
\colhead{(\kms)}        &
\colhead{}              \\
\colhead{(1)}           &
\colhead{(2)}           &
\colhead{(3)}           &
\colhead{(4)}           &
\colhead{(5)}           
}
\startdata
\multicolumn{5}{c}{$z$ = 1.78169 (HE~0141$-$3932)}                               \\
\hline
\ion{H }{1} & $-$82      &  $>$14.36$^b$             &  42.3 $\pm$  0.3      & x \\
            &    11      &  $>$15.28$^b$             &  21.5 $\pm$  0.5      &   \\
\ion{Mg}{2} &     0      &     12.05 $\pm$ 0.01      &   4.1 $\pm$  0.2      & x \\
\ion{Fe}{2} &     2      &     11.29 $\pm$ 9.99      &   1.1 $\pm$ 99.9      &   \\
\ion{Si}{2} &     0      &     12.26 $\pm$ 0.01      &   4.5 $\pm$  0.3      &   \\
\ion{Al}{2} &     0      &     11.16 $\pm$ 0.07      &   5.6 $\pm$  1.5      &   \\
\ion{C }{2} &     0      &     13.16 $\pm$ 0.01      &   5.6 $\pm$  0.3      &   \\
            &    28      &     12.33 $\pm$ 0.08      &   6.4 $\pm$  1.9      &   \\
\ion{Al}{3} &     0      &     11.51 $\pm$ 0.04      &   3.4 $\pm$  0.9      &   \\
\ion{Si}{4} &     0      &     12.46 $\pm$ 0.02      &   6.0 $\pm$  0.4      &   \\
            &    29      &     12.11 $\pm$ 0.04      &   5.8 $\pm$  0.9      &   \\
\ion{C }{4} & $-$77      &     12.57 $\pm$ 0.03      &   8.7 $\pm$  0.9      & x \\
            &     1      &     12.80 $\pm$ 0.02      &   7.4 $\pm$  0.4      & x \\
            &    31      &     13.13 $\pm$ 0.01      &   9.0 $\pm$  0.3      & x \\
\hline 
\multicolumn{5}{c}{$z$ = 1.68079 (HE~0429$-$4091)}                               \\
\hline
\ion{H }{1} &    14      &  $>$15.57$^b$             &  48.3 $\pm$  2.3      &   \\
            &   (64)$^c$ & ($>$14.6)$^c$             & (34)$^c$              & x \\
\ion{Mg}{2} &     0      &     11.52 $\pm$ 0.09      &   6.9 $\pm$  2.3      & x \\
\ion{Si}{2} &     3      &     11.81 $\pm$ 0.12      &   3.7 $\pm$  2.7      &   \\
\ion{Al}{2} &     3      &     10.80 $\pm$ 0.15      &   4.0 $\pm$  3.2      &   \\
\ion{C }{2} &     1      &     12.78 $\pm$ 0.04      &   5.6 $\pm$  0.9      &   \\
\ion{Si}{3} & $-$31      &     12.62 $\pm$ 0.09      &  19.9 $\pm$  5.8      &   \\
            &     2      &     12.87 $\pm$ 0.09      &   7.9 $\pm$  1.3      &   \\
\ion{Si}{4} & $-$32      &     12.73 $\pm$ 0.02      &  20.1 $\pm$  1.4      &   \\
            &     2      &     13.28 $\pm$ 0.01      &   5.4 $\pm$  0.1      &   \\
\ion{C }{4} & $-$47      &     13.25 $\pm$ 0.16      &  10.0 $\pm$  1.4      &   \\
            & $-$31      &     13.24 $\pm$ 0.41      &  10.6 $\pm$  3.5      &   \\
            & $-$12      &     13.40 $\pm$ 0.28      &  23.6 $\pm$ 11.8      &   \\
            &     3      &     14.04 $\pm$ 0.02      &   6.1 $\pm$  0.2      &   \\
            &($-$47)$^c$ &    (13.27 $\pm$ 0.03)$^c$ & (10.5 $\pm$  0.3)$^c$ & x \\
            &($-$28)$^c$ &    (13.48 $\pm$ 0.02)$^c$ & (13.8 $\pm$  0.5)$^c$ & x \\
            &    (5)$^c$ &    (13.85 $\pm$ 0.01)$^c$ &  (5.3 $\pm$  0.1)$^c$ & x \\
\hline
\multicolumn{5}{c}{$z$ = 1.75570 (HE~2243$-$6031)}                               \\
\hline
\ion{H }{1} &  $-$1      &  $>$14.85$^b$             &  18.6 $\pm$  1.4      &   \\
\ion{Mg}{2} &     0      &     12.64 $\pm$ 0.01      &   5.4 $\pm$  0.1      & x \\
\ion{Al}{3} &     2      &     11.89 $\pm$ 0.06      &   4.7 $\pm$  1.3      &   \\
\ion{Si}{3} &     0      &     13.08 $\pm$ 0.07      &   9.8 $\pm$  0.8      &   \\
\ion{Si}{4} &     1      &     13.65 $\pm$ 0.01      &   6.8 $\pm$  0.1      &   \\
\ion{C }{4} &  $-$1      &     14.72 $\pm$ 0.06      &   9.9 $\pm$  0.3      &   \\
            & ($-$1)$^c$ &    (14.73 $\pm$ 0.06)$^c$ &  (9.8 $\pm$  0.3)$^c$ & x \\
\enddata 
\tablenotetext{a}{Measured column density of cloud is used to optimize
                  on this transition in our photoionization model.}
\tablenotetext{b}{Column densities of \ion{H}{1} components measured
                  from the best model are listed in column~9 of
                  Table~\ref{tab6}.}
\tablenotetext{c}{Voigt profile fit after removing contributions to
                  this ion from the other phases.}
\end{deluxetable}
\clearpage

\begin{deluxetable}{lccll}
\tabletypesize{\scriptsize}
\tablecaption{Model Constraints for Three Weak MgII Systems \label{tab4}}
\tablewidth{0pt}
\tablehead{
\colhead{Cloud}         &
\colhead{parameter}     &
\colhead{constraint}    &
\colhead{line}          &
\colhead{condition}     \\
\colhead{(1)}           &
\colhead{(2)}           &
\colhead{(3)}           &
\colhead{(4)}           &
\colhead{(5)}           
}
\startdata
\hline
\multicolumn{5}{c}{System~1} \\
\hline
\ion{Mg}{2}~(0~\kms)   &  $\log U$  &  $>$ $-$3.8       & \ion{Fe}{2}                            & to avoid over-production \\
                       &            &  $<$ $-$3.7       & \ion{Al}{2} and \ion{Si}{2}            & to avoid over-production \\
                       &            & ($>$ $-$6.0       & \ion{Mg}{1}                            & to avoid over-production) \\
                       &            & ($<$ $-$2.8       & \ion{C}{4} and \ion{Si}{4}             & to avoid over-production) \\
                       &  $\log Z$  &  $>$ $-$0.7       & \lya                                   & to avoid over-production \\
                       &            &  $<$ $-$0.5       & \ion{Fe}{2}                            & to avoid over-production at $\log U$ $\sim$ $-$3.7 \\
\ion{C}{4}~($-$77~\kms)&  $\log U$  &  $>$ $-$1.5       & \ion{Si}{4} and LI$^a$                 & to avoid over-production \\
                       &            &  $<$ $-$1.0       & \ion{N}{5}                             & to avoid over-production \\
                       &  $\log Z$  &  $>$ $-$1.8       & \lya                                   & to avoid over-production \\
                       &            &  $<$    0.5       & LI$^a$                                 & to avoid over-production \\
\ion{C}{4}~(1~\kms)    &  $\log U$  &  $>$ $-$2.35      & \ion{Si}{4} and LI$^a$                 & to avoid over-production \\
                       &            &  $<$ $-$2.25      & \ion{N}{5}                             & to avoid over-production \\
                       &  $\log Z$  &  $>$ $-$0.7       & \lya                                   & to avoid over-production \\
                       &            &  $<$    0.1       & LI$^a$                                 & to avoid over-production \\
\ion{C}{4}~(31~\kms)   &  $\log U$  &  $>$ $-$1.9       & \ion{Si}{4} and LI$^a$                 & to avoid over-production \\
                       &            &  $<$ $-$1.8       & \ion{N}{5}                             & to avoid over-production \\
                       &  $\log Z$  &  $-$0.5 -- $-$0.4 & \lya                                   & to match the observed profile \\
                       &            & ($>$ $-$0.6       & \lya                                   & to avoid over-production) \\
\lya~($-$82~\kms)      &  $\log Z$  &  $<$    0.0       &                                        & to produce no other transitions if $\log U$ $\sim$ $-$3.0 \\
                       &            &  $<$ $-$2.0       &                                        & to produce no other transitions if $\log U$ $\sim$ $-$1.5  \\
\hline
\multicolumn{5}{c}{System~2} \\
\hline
\ion{Mg}{2}~(0~\kms)   &  $\log U$  &  $\sim$ $-$2.0    & \ion{C}{4} and \ion{Si}{4}             & to match the observed profiles \\
                       &            & ($>$ $-$3.5       & \ion{Fe}{2}                            & to avoid over-production) \\
                       &            & ($<$ $-$2.0       & \ion{C}{4}                             & to avoid over-production) \\
                       &  $\log Z$  &  $>$ $-$0.2       & \ion{Al}{2}                            & to avoid over-production \\ 
                       &            &  $<$    0.1       & \ion{C}{4} and \ion{Si}{4}             & to avoid under-production  \\
                       &            & ($>$ $-$3.0       & \lya                                   & to avoid over-absorption of the blue-wing) \\
\ion{C}{4}~($-$47~\kms)&  $\log U$  &  $-$1.7 -- $-$1.6 & \ion{Si}{3} and \ion{Si}{4}            & to match the observed profile \\
                       &  $\log Z$  &  $-$1.1 -- $-$1.0 & \lya                                   & to account for the blueward edge of \lya\ profile \\
\ion{C}{4}~($-$28~\kms)&  $\log U$  &  $\sim$ $-$1.8    & \ion{Si}{3} and \ion{Si}{4}            & to match the observed profile \\
                       &  $\log Z$  &  $>$ $-$1.8       & \lya                                   & to avoid over-production \\
\ion{C}{4}~(5~\kms)    &  $\log U$  &  $\sim$ $-$1.8    & \ion{Si}{3} and \ion{Si}{4}            & to match the observed profile \\
                       &  $\log Z$  &  $>$ $-$2.9       & \lya                                   & to avoid over-production \\
                       &            &  $<$ $-$0.3       & LI$^a$                                 & to avoid over-production \\
\lya~(64~\kms)         &  $\log Z$  &  $<$ $-$2.0       & \ion{C}{4} and \ion{N}{5}              & to avoid over-production if $\log U$ $>$ $-$2.0 \\
                       &            &  $<$    0.0       & LI$^a$                                 & to avoid over-production if $\log U$ $<$ $-$2.5 \\
\hline
\multicolumn{5}{c}{System~3 (Model~1)} \\
\hline
\ion{Mg}{2}~(0~\kms)   &  $\log U$  &  $\sim$ $-$2.0    & \ion{Al}{2}, \ion{Al}{3}, \ion{Si}{2}  & to match the observed profile \\
                       &            & ($>$ $-$3.5       & \ion{Fe}{2}                            & to avoid over-production) \\
                       &            & ($<$ $-$2.0       & \ion{Si}{4}                            & to avoid over-production) \\
                       &            & ($<$ $-$1.5       & \ion{C}{4}                             & to avoid over-production) \\
                       &  $\log Z$  &  $>$    0.9       & \ion{Al}{3}                            & to avoid over-production \\
                       &            & ($>$    0.4       & \ion{Si}{4}                            & to avoid over-production) \\
                       &            & ($>$    0.0       & \lya                                   & to avoid over-production) \\
\ion{C}{4}~(0~\kms)    &  $\log U$  &  $-$1.5 -- $-$1.4 & \ion{Si}{4}                            & to match the observed profile \\
                       &  $\log Z$  &  $-$0.1 -- 0.0    & \lya                                   & to match the observed profile \\
\hline
\multicolumn{5}{c}{System~3 (Model~2)} \\
\hline
\ion{Mg}{2}~(0~\kms)   &  $\log U$  &  $\sim$ $-$2.0    & \ion{Al}{2}, \ion{Al}{3}, \ion{Si}{2}  & to minimize over-production$^b$ \\
                       &            & ($>$ $-$3.5       & \ion{Fe}{2}                            & to avoid over-production) \\
                       &            & ($<$ $-$2.0       & \ion{Si}{4}                            & to avoid over-production) \\
                       &            & ($<$ $-$1.5       & \ion{C}{4}                             & to avoid over-production) \\
                       &  $\log Z$  &  $>$    0.0       & \lya                                   & to avoid over-production$^c$ \\
\ion{C}{4}~(0~\kms)    &  $\log U$  &  $\sim$ $-$1.1    & \ion{Si}{4}                            & to match the observed profile \\
                       &  $\log Z$  &  $>$ $-$0.2       & \lya                                   & to avoid over-production \\
\enddata 
\tablenotetext{a}{Low and intermediate--ionization transitions
                  including \ion{Mg}{1}, \ion{Mg}{2}, \ion{O}{1},
                  \ion{C}{2}, \ion{Fe}{2}, \ion{Si}{2}, \ion{Si}{3},
                  \ion{Al}{2}, \ion{Al}{3}.}
\tablenotetext{b}{To match these profiles a reduction of the aluminum
                  abundance by 0.5 and the silicon abundance by 0.2
                  can be applied.}
\tablenotetext{c}{A $\log Z$ $>$ 0.0 is not required by \ion{Al}{2},
                  \ion{Al}{3}, and \ion{Si}{2} if an abundance pattern
                  adjustment is used.}
\end{deluxetable}
\clearpage

\begin{deluxetable}{lcccccccc}
\tabletypesize{\scriptsize}
\setlength{\tabcolsep}{0.04in}
\tablecaption{Acceptable Models for Three Weak MgII Systems \label{tab5}}
\tablewidth{0pt}
\tablehead{
\colhead{$z_{abs}$}                  &
\colhead{ion}                        &
\colhead{\delv}                      &
\colhead{$\log U$}                   &
\colhead{$\log (Z/Z_{\odot})$}       &
\colhead{$\log (Z/Z_{\odot})_{min}$} &
\colhead{Size}                       &
\colhead{$b$}                        &
\colhead{$\log N$(\ion{H}{1})}       \\
\colhead{}                    &
\colhead{}                    &
\colhead{(\kms)}              &
\colhead{}                    &
\colhead{}                    &
\colhead{}                    &
\colhead{(kpc)}               &
\colhead{(\kms)}              &
\colhead{(cm$^{-2}$)}         \\
\colhead{(1)}                 &
\colhead{(2)}                 &
\colhead{(3)}                 &
\colhead{(4)}                 &
\colhead{(5)}                 &
\colhead{(6)}                 &
\colhead{(7)}                 &
\colhead{(8)}                 &
\colhead{(9)}                 
}
\startdata
 1.78169 ........................ & \ion{Mg}{2} &     0 & $-$3.8 to $-$3.7 & $-$0.7 to $-$0.5     & $>$ $-$0.8     & 0.00051 to 0.0012 &  4 & 15.81 to 15.96 \\
                                  & \ion{C}{4}  & $-$77 & $-$1.5 to $-$1.0 & $-$1.8 to    0.5     & $>$ $-$1.9     & 0.0072  to 13.    &  9 & 12.58 to 13.42 \\
                                  & \ion{C}{4}  &     1 & $\sim$$-$2.3     & $-$0.7 to    0.1     & $>$ $-$0.8     & 0.016   to 0.083  &  7 & 14.46 to 14.57 \\
                                  & \ion{C}{4}  &    31 & $-$1.9 to $-$1.8 & $-$0.5 to $-$0.4     & $>$ $-$0.6     & 0.11    to 0.15   &  9 & 14.11 to 14.45 \\
                                  & \lya        & $-$82 & ...              &       $<$    0.0     & $<$    0.0     & ...               & 42 & ...            \\
 1.68079 ........................ & \ion{Mg}{2} &     0 & $\sim$$-$2.0     & $-$0.2 to    0.1     & $>$ $-$3.0     & 0.085   to 0.23   &  7 & 14.51 to 14.83 \\
                                  & \ion{C}{4}  & $-$47 & $-$1.7 to $-$1.6 & $-$1.1 to $-$1.0     & $>$ $-$1.1     & 0.78    to 1.2    & 11 & 14.46 to 14.68 \\
                                  & \ion{C}{4}  & $-$28 & $\sim$$-$1.8     &       $>$ $-$1.8     & $>$ $-$1.8     & $<$11.            & 14 & $<$15.83       \\
                                  & \ion{C}{4}  &     5 & $\sim$$-$1.8     & $-$2.9 to $-$0.3     & $>$ $-$3.0     & 0.54    to 360    &  5 & 14.76 to 17.36 \\
                                  & \lya        &    64 & ...              &       $<$    0.0     & $<$    0.0     & ...               & 34 & ...            \\
 1.75570 (Model~1) .......        & \ion{Mg}{2} &     0 & $\sim$$-$2.0     &       $>$    0.9     & $>$    0.0     & $<$0.019          &  5 & $<$ 14.70      \\
                                  & \ion{C}{4}  &     0 & $-$1.5 to $-$1.4 & $-$0.1 to    0.0     & $>$ $-$0.2     & 3.1     to 5.0    & 10 & 14.94 to 15.00 \\
 1.75570 (Model~2) .......        & \ion{Mg}{2} &     0 & $\sim$$-$2.0     &       $\sim$ 0.0     & $>$    0.0     & $\sim$1.5         &  5 & $\sim$15.72    \\
                                  & \ion{C}{4}  &     0 & $\sim$$-$1.1     &       $>$ $-$0.2     & $>$ $-$0.2     & $<$17.            & 10 & $<$ 14.81      \\
\enddata 
\end{deluxetable}
\clearpage

\pagestyle{empty}
\begin{deluxetable}{cccccccccccccccc}
\rotate
\tabletypesize{\scriptsize}
\setlength{\tabcolsep}{0.015in}
\tablecaption{Best Fit Models for Three Weak MgII Systems \label{tab6}}
\tablewidth{0pt}
\tablehead{
\colhead{Cloud}                 &
\colhead{$v$}                   &
\colhead{$\log (Z/Z_{\odot})$}  &
\colhead{$\log U$}              &
\colhead{$n_{H}$}               &
\colhead{Size}                  &
\colhead{$T$}                   &
\colhead{$\log N_{tot}$(H)}     &
\colhead{$\log N$(\ion{H}{1})}  &
\colhead{$\log N$(\ion{Mg}{2})} &
\colhead{$\log N$(\ion{Si}{4})} &
\colhead{$\log N$(\ion{C}{4})}  &
\colhead{$b$(H)}                &
\colhead{$b$(Mg)}               &
\colhead{$b$(C)}                &
\colhead{Abundance Pattern}     \\
\colhead{}                 &
\colhead{(\kms)}           &
\colhead{}                 &
\colhead{}                 &
\colhead{(cm$^{-3}$)}      &
\colhead{(kpc)}            &
\colhead{(K)}              &
\colhead{(cm$^{-2}$)}      &
\colhead{(cm$^{-2}$)}      &
\colhead{(cm$^{-2}$)}      &
\colhead{(cm$^{-2}$)}      &
\colhead{(cm$^{-2}$)}      &
\colhead{(\kms)}           &
\colhead{(\kms)}           &
\colhead{(\kms)}           &
\colhead{}                 \\
\colhead{(1)}              &
\colhead{(2)}              &
\colhead{(3)}              &
\colhead{(4)}              &
\colhead{(5)}              &
\colhead{(6)}              &
\colhead{(7)}              &
\colhead{(8)}              &
\colhead{(9)}              &
\colhead{(10)}             &
\colhead{(11)}             &
\colhead{(12)}             &
\colhead{(13)}             &
\colhead{(14)}             &
\colhead{(15)}             &
\colhead{(16)}             
}
\startdata
\multicolumn{16}{c}{$z$ = 1.78169 (HE~0141$-$3932)} \\
\hline
 \ion{Mg}{2} 1 &     0 &    $-$0.6 &    $-$3.7 & 0.095  & 0.00096& 11100 & 17.45 & 15.87 & 12.06 & 10.27 & 10.10 & 14 &  4 &  5 & solar \\
 \ion{C}{4}  1 & $-$77 &    $-$0.5 &    $-$1.0 & 0.00019& 0.43   & 24400 & 17.40 & 12.94 &  8.13 & 10.78 & 12.58 & 21 &  7 &  9 & solar \\
 \ion{C}{4}  2 &     1 &       0.0 &    $-$2.3 & 0.0038 & 0.019  &  8730 & 17.35 & 14.49 & 11.46 & 12.35 & 12.81 & 14 &  7 &  7 & solar \\
 \ion{C}{4}  3 &    31 &    $-$0.5 &    $-$1.8 & 0.0012 & 0.15   & 17300 & 17.75 & 14.19 & 10.45 & 12.14 & 13.12 & 18 &  8 &  9 & solar \\
 \lya        1 & $-$82 &    $-$1.0 &    $-$2.5 & 0.0060 & 0.0094 & 17300 & 17.24 & 14.36 & 10.32 & 11.00 & 11.55 & 42 & 39 & 39 & solar \\
\hline
\multicolumn{16}{c}{$z$ = 1.68079 (HE~0429$-$4091)} \\
\hline
 \ion{Mg}{2} 1 &     0 &    $-$0.1 &    $-$2.0 & 0.0018 & 0.17   & 11200 & 17.98 & 14.74 & 11.53 & 12.94 & 13.63 & 15 &  7 &  7 & solar \\
 \ion{C}{4}  1 & $-$47 &    $-$1.0 &    $-$1.6 & 0.00072& 1.2    & 24500 & 18.43 & 14.55 & 10.12 & 12.02 & 13.26 & 22 & 10 & 11 & solar \\
 \ion{C}{4}  2 & $-$28 &    $-$1.0 &    $-$1.8 & 0.0012 & 1.4    & 22600 & 18.70 & 15.04 & 10.78 & 12.45 & 13.49 & 23 & 13 & 14 & solar \\
 \ion{C}{4}  3 &     5 &    $-$1.0 &    $-$1.8 & 0.0011 & 3.4    & 22500 & 19.08 & 15.42 & 11.16 & 12.83 & 13.86 & 18 &  4 &  5 & $-$0.5dex (N)\\
 \lya        1 &    64 &    $-$1.0 &    $-$2.5 & 0.0057 & 0.016  & 17300 & 17.45 & 14.57 & 10.53 & 11.20 & 11.75 & 34 & 30 & 30 & solar \\
\hline 
\multicolumn{16}{c}{$z$ = 1.75570 (HE~2243$-$6031)} \\
\hline 
 Model~1       &       &           &           &        &        &       &       &       &       &       &       &    &    &    &       \\
 \ion{Mg}{2} 1 &     0 &       1.0 &    $-$2.0 & 0.0015 & 0.023  &   424 & 17.02 & 14.62 & 12.64 & 13.11 & 13.59 &  6 &  6 &  6 & solar \\
 \ion{C}{4}  1 &     0 &    $-$0.1 &    $-$1.4 & 0.00047& 5.0    & 15200 & 18.86 & 14.94 & 11.27 & 13.42 & 14.73 & 18 &  9 & 10 & solar \\
\hline
 Model~2       &       &           &           &        &        &       &       &       &       &       &       &    &    &    &       \\
 \ion{Mg}{2} 1 &     0 &       0.0 &    $-$2.0 & 0.0018 & 1.5    &  9870 & 18.93 & 15.72 & 12.63 & 13.81 & 14.65 & 14 &  6 &  6 & $-$0.5dex(Al), $-$0.2dex(Si) \\
 \ion{C}{4}  1 &     0 &       0.2 &    $-$1.1 & 0.00024& 5.8    & 14700 & 18.63 & 14.43 & 10.61 & 13.14 & 14.72 & 18 & 10 & 10 & solar \\
\enddata 
\end{deluxetable}
\clearpage

\begin{deluxetable}{ccccccccccccc}
\rotate
\tabletypesize{\scriptsize}
\tablecaption{Single-Cloud Weak MgII Systems from Literature \label{tab7}}
\tablewidth{0pt}
\tablehead{
\colhead{QSO}             &
\colhead{\zabs}           &
\colhead{$\log N_{MgII}$} &
\colhead{$b_{MgII}$}      &
\colhead{$\log U$}        &
\colhead{$\log Z$}        &
\colhead{$\log n_{H}$}    &
\colhead{$\log d$}        &
\colhead{$EW$(2796)}      &
\colhead{$T_{gas}$}       &
\colhead{$\log N_{HI}$}   &
\colhead{$\log N_{tot}$}  &
\colhead{ref.$^a$}        \\
\colhead{}                &
\colhead{}                &
\colhead{(\cmm)}          &
\colhead{(\kms)}          &
\colhead{}                &
\colhead{}                &
\colhead{(\cmmm)}         &
\colhead{(pc)}            &
\colhead{(\AA)}           &
\colhead{(K)}             &
\colhead{(\cmm)}          &
\colhead{(\cmm)}          &
\colhead{}                \\
\colhead{(1)}             &
\colhead{(2)}             &
\colhead{(3)}             &
\colhead{(4)}             &
\colhead{(5)}             &
\colhead{(6)}             &
\colhead{(7)}             &
\colhead{(8)}             &
\colhead{(9)}             &
\colhead{(10)}            &
\colhead{(11)}            &
\colhead{(12)}            &
\colhead{(13)}            
}
\startdata
 Q~1421$+$331   & 0.4564 & 13.07$\pm$0.06 &  7.7$\pm$0.6 &         ...         &        ...        &        ...        &        ...        & 0.179$\pm$0.019 &  ...  &      ...     & ...  & 1 \\
 Q~1354$+$193   & 0.5215 & 11.91$\pm$0.05 &  4.9$\pm$0.9 &         ...         & $>$ $-$1.5        &        ...        &        ...        & 0.030$\pm$0.007 &  ...  &      ...     & ...  & 1 \\
 Q~0002$+$051   & 0.5915 & 12.63$\pm$0.01 &  6.8$\pm$0.2 & $>$ $-$3.5          &        ...        &        $<$ $-$2.1 &    1.2 --     4.5 & 0.103$\pm$0.008 &  ...  &      ...     & ...  & 1 \\
 Q~0454$+$036   & 0.6428 & 12.74$\pm$0.02 &  5.8$\pm$0.3 & $-$4.5  --  $-$4.2  & $-$1.0 --     0.0 & $-$1.4 --  $-$1.1 &    0.3 --     0.9 & 0.118$\pm$0.008 &  ...  &      ...     & ...  & 1 \\
 PG~1634$+$706  & 0.6534 & 11.80$\pm$0.10 &  4.0$\pm$2.0 & $-$4.0  --  $-$3.0  & $-$1.5 --     0.0 & $-$1.2 --  $-$1.2 & $-$0.7 --     2.0 & 0.031$\pm$0.006 & 11000 & 15.1 -- 16.7 & 17.5 & 4 \\
 Q~0823$-$223   & 0.7055 & 12.40$\pm$0.02 & 13.3$\pm$0.6 & $-$3.6  --  $-$2.4  &        ...        & $-$3.2 --  $-$2.0 &    0.8 --     4.7 & 0.092$\pm$0.007 &  ...  &      ...     & ...  & 1 \\
 PG~1634$+$706  & 0.8181 & 12.04$\pm$0.03 &  2.1$\pm$0.4 & $-$6.0  --  $-$4.0  & $>$ 0.3           & $-$1.2 --  $-$1.2 & $-$3.0 --  $-$1.0 & 0.030$\pm$0.005 &  4600 & 15.6 -- 15.9 & 16.3 & 4 \\
 Q~1421$+$331   & 0.8433 & 13.10$\pm$0.10 &  3.2$\pm$0.2 & $-$5.0  --  $-$3.0  &        ...        & $-$2.5 --  $-$0.8 &    0   --     1.1 & 0.086$\pm$0.008 &  ...  &      ...     & ...  & 1 \\
 Q~0002$+$051   & 0.8665 & 11.89$\pm$0.04 &  2.7$\pm$0.8 & $>$ $-$3.6          & $>$ $-$1.0        &        $<$ $-$1.8 &    0.5 --     3.5 & 0.023$\pm$0.008 &  ...  &      ...     & ...  & 1 \\
 PG~1241$+$176  & 0.8954 & 11.70$\pm$0.00 &  6.8$\pm$0.0 & $-$4.0  --  $-$4.0  & $>$ $-$1.7        & $-$1.4 --  $-$1.4 &    2.3 --     2.3 & 0.018$\pm$0.005 &  6500 & 15.2 -- 15.2 & 16.3 & 2 \\
 PG~1634$+$706  & 0.9056 & 12.47$\pm$0.01 &  2.8$\pm$0.1 & $-$3.0  --  $-$2.7  & $>$ 0.0           & $-$2.5 --  $-$2.5 &    1.5 --     2.0 & 0.034$\pm$0.002 &  9000 & 15.7 -- 15.7 & 18.1 & 4 \\
 Q~0454$+$036   & 0.9315 & 12.29$\pm$0.08 &  1.5$\pm$0.2 & $-$4.7  --  $-$3.8  & $-$1.0 --     0.0 & $-$1.4 --  $-$0.5 &        $<$    0.3 & 0.042$\pm$0.005 &  ...  &      ...     & ...  & 1 \\
 Q~1206$+$456   & 0.9343 & 12.05$\pm$0.02 &  7.5$\pm$0.5 & $-$3.7  --  $-$1.7  & $>$ $-$1.0        & $-$3.5 --  $-$1.5 &    1   --     4.0 & 0.049$\pm$0.005 &  ...  &      ...     & ...  & 1 \\
 Q~0002$+$051   & 0.9560 & 12.15$\pm$0.02 &  7.5$\pm$0.6 & $>$ $-$3.7          & $>$ $-$1.0        &        $<$ $-$1.6 &    $>$ 0          & 0.052$\pm$0.007 &  ...  &      ...     & ...  & 1 \\
 Q~1213$-$003   & 1.1278 & 12.11$\pm$0.05 &  1.9$\pm$0.4 & $>$ $-$4.6          &        ...        &        $<$ $-$0.6 &        ...        & 0.036$\pm$0.006 &  ...  &      ...     & ...  & 1 \\
 Q~0958$+$551   & 1.2113 & 12.41$\pm$0.03 &  3.3$\pm$0.3 & $>$ $-$3.5          &        ...        &        $<$ $-$1.7 &        ...        & 0.060$\pm$0.007 &  ...  &      ...     & ...  & 1 \\
 Q~0958$+$551   & 1.2724 & 12.57$\pm$0.02 &  3.9$\pm$0.2 & $-$3.7  --  $-$2.8  & $>$ $-$2.5        & $-$2.4 --  $-$1.5 &    1   --     4.2 & 0.081$\pm$0.007 &  ...  &      ...     & ...  & 1 \\
 HE~2347$-$4342 & 1.4054 & 11.87$\pm$0.01 &  7.4$\pm$0.1 & $-$3.0  --  $-$2.5  &        ...        & $-$2.3 --  $-$1.8 &        ...        & 0.076$\pm$0.001 &  ...  &      ...     & ...  & 3 \\
 Q~0002$-$422   & 1.4465 & 12.09$\pm$0.01 &  6.0$\pm$0.1 & $-$3.5  --  $-$3.0  &        ...        & $-$1.8 --  $-$1.3 &        ...        & 0.042$\pm$0.001 &  ...  &      ...     & ...  & 3 \\
 HE~0001$-$2340 & 1.6515 & 12.56$\pm$0.01 &  2.9$\pm$0.0 & $-$4.0  --  $-$3.5  & $>$ $-$1.5        &        $<$ $-$1.3 &        $<$    1.7 & 0.070$\pm$0.001 &  ...  &      ...     & ...  & 3 \\
 HE~0429$-$4901 & 1.6808 & 11.52$\pm$0.09 &  6.9$\pm$2.3 & $-$2.05 --  $-$1.95 & $-$0.2 --     0.1 & $-$2.8 --  $-$2.7 &    1.9 --     2.4 & 0.015$\pm$0.001 & 11200 & 14.5 -- 14.8 & 18.0 & 5 \\
 HE~0151$-$4326 & 1.7085 & 11.86$\pm$0.01 &  3.9$\pm$0.1 & $-$4.0  --  $-$2.5  & $>$ $-$1.5        &        $<$ $-$0.7 &        $<$    3.1 & 0.027$\pm$0.001 &  ...  &      ...     & ...  & 3 \\
 HE~2243$-$6031 & 1.7557 & 12.64$\pm$0.01 &  5.4$\pm$0.1 & $-$2.05 --  $-$1.95 & $>$ 0.9           & $-$2.8 --  $-$2.7 &        $<$    1.3 & 0.121$\pm$0.003 &   460 &     $<$ 14.7 & 17.0 & 5 \\  
 HE~0141$-$3932 & 1.7817 & 12.05$\pm$0.01 &  4.1$\pm$0.2 & $-$3.8  --  $-$3.7  & $-$0.7 --  $-$0.5 & $-$1.0 --  $-$0.9 & $-$0.3 --     0.1 & 0.039$\pm$0.001 & 11100 & 15.8 -- 16.0 & 17.5 & 5 \\
 HE~2347$-$4342 & 1.7962 & 13.26$\pm$0.02 &  4.2$\pm$0.1 & $-$4.0  --  $-$3.2  & $>$ $-$1.0        &        $<$ $-$0.7 &        $<$    2.4 & 0.146$\pm$0.001 &  ...  &      ...     & ...  & 3 \\
 HE~0940$-$1050 & 2.1745 & 11.91$\pm$0.01 &  4.6$\pm$0.2 & $-$3.7  --  $-$2.5  & $>$ $-$2.0        &        $<$ $-$2.2 &        $<$    3.5 & 0.028$\pm$0.001 &  ...  &      ...     & ...  & 3 
\enddata 
\tablenotetext{a}{
(1) \citet{rig02} and \citet{chu99},
(2) \citet{din05},
(3) \citet{lyn07},
(4) \citet{cha03},
(5) This paper.
}
\end{deluxetable}
\clearpage
\pagestyle{plaintop}


\clearpage

\clearpage

\clearpage

\clearpage

\clearpage

\clearpage

\clearpage

\end{document}